# Extreme $^{54}$Cr-rich nano-oxides in the CI chondrite Orgueil -Implication for a late supernova injection into the Solar System

L. Qin$^{1}$*$^{a}$, L. R. Nittler$^{1}$, C. M. O'D. Alexander$^{1}$, J. Wang$^{1}$, F. J. Stadermann$^{2}$, and R. W. Carlson$^{1}$

$^{1}$Department of Terrestrial Magnetism, Carnegie Institution of Washington, 5241 Broad Branch RD. NW, Washington, DC 20015, USA.

$^{2}$Laboratory for Space Sciences and Department of Physics, Washington University, One Brookings Drive, St. Louis, Mo 63130, USA.

$^{a}$Current address: Center for Isotope Geochemistry, Lawrence Berkeley National Laboratory, 1 Cyclotron RD., MS 70A4418, Berkeley, CA 94720.

*Corresponding Author: e-mail: lqin@lbl.gov

**Abstract** Systematic variations in $^{54}$Cr/$^{52}$Cr ratios between meteorite classes (Qin et al., 2010a; Trinquier et al., 2007) point to large scale spatial and/or temporal isotopic heterogeneity in the solar protoplanetary disk. Two explanations for these variations have been proposed, with important implications for the formation of the Solar System: heterogeneous seeding of the disk with dust from a supernova, or energetic-particle irradiation of dust in the disk. The key to differentiating between them is identification of the carrier(s) of the $^{54}$Cr anomalies. Here we report the results of our recent NanoSIMS imaging search for the $^{54}$Cr-rich carrier in the acid-resistant residue of the CI chondrite Orgueil. A total of 10 regions with extreme $^{54}$Cr-excesses ($\delta^{54}$Cr values up to 1500 ‰) were found. Comparison between SEM, Auger and NanoSIMS analyses showed that these $^{54}$Cr-rich regions are associated with one or more sub-micron (typically less than 200 nm) Cr oxide grains, most likely spinels. Because the size of the NanoSIMS primary O$^-$ ion beam is larger than the typical grain size on the sample mount, the measured anomalies are lower limits, and we estimate that the actual $^{54}$Cr enrichments in three grains are at least 11 times Solar and in one of these may be as high as 50 times Solar. Such compositions strongly favor a Type II supernova origin. The variability in bulk $^{54}$Cr/$^{52}$Cr between meteorite classes argues for a heterogeneous distribution of the $^{54}$Cr carrier in the solar protoplanetary disk following a late supernova injection event. Such a scenario is also supported by the O-isotopic distribution and variable abundances in different planetary materials of other presolar oxide and silicate grains from supernovae.

## 1. Introduction

Planetary-scale $^{54}$Cr isotope anomalies have been well documented in the inner Solar System, as well as amongst primitive and differentiated meteorites (Birck and Allègre, 1984; Papanastassiou, 1986; Podosek et al., 1997; Qin et al., 2010a; Rotaru et al., 1992; Shukolyukov and Lugmair, 2006; Trinquier et al., 2007), and are systematic enough to be used as a classification tool. The only isotope system with a comparable systematic variation is O (Clayton, 2003). The $^{54}$Cr variations between different meteorite classes have been attributed to the spatial and/or temporal heterogeneity of a $^{54}$Cr-rich component in the solar nebula that either formed in a supernova (Qin et al., 2010a; Shukolyukov and Lugmair, 2006; Trinquier et al., 2007) or through energetic particle irradiation of Fe/Ni-rich grains in the Solar System (Qin et al., 2010a). Determining the chemical and isotopic composition of the $^{54}$Cr carrier phase(s) would provide insights into the processes that governed such large-scale isotopic heterogeneity and the evolution in the solar nebula.

The identity of the $^{54}$Cr carrier phase(s) is a long-standing problem (Podosek et al., 1997; Rotaru et al., 1992; Trinquier et al., 2007). Until recently, a few refractory calcium-aluminum-rich inclusions (CAIs) were the only chondritic components known to have large $^{54}$Cr excesses (up to 5 ‰) (Birck and Allègre, 1984; Papanastassiou, 1986). Although CAIs may represent an isotopically anomalous reservoir, their low abundance and low Cr contents means that they are not the major source of anomalous $^{54}$Cr in bulk meteorites.

Previous work has shown that the $^{54}$Cr-rich phase(s) can be concentrated in the acid-resistant residues of chondrites (Alexander, 2002; Alexander and Carlson, 2007; Qin et al., 2010a). The carrier is at least partly soluble in hot HCl, with the leachates showing maximum $\delta^{54}Cr/^{52}Cr$ excesses of ~20 ‰, accompanied by $\delta^{53}Cr/^{52}Cr$ of $\geq$ -0.2 ‰ (Podosek et al., 1997; Qin et al., 2010a; Rotaru et al., 1992). The $^{54}$Cr anomalies in the bulk residues are several fold lower than that in the HCl leachates, indicating dilution by isotopically normal or light Cr in the residues (Qin et al., 2010a). Mass balance calculations show that most of the anomalous Cr is unleachable in HCl (Qin et al., 2010a), suggesting that there are either multiple carrier phases or, perhaps, that the carrier grains are only partially leachable (e.g., due to radiation damage).

Firmer constraints on the origin of the $^{54}$Cr-rich carriers, the magnitude of the $^{54}$Cr anomalies they carry, and their chemistry and mineralogy must rely on in-situ isotopic and elemental studies of the residues. In the work described here, we have conducted a successful search for the carrier phase of the $^{54}$Cr excess in grains from an acid-resistant residue of the CI chondrite Orgueil using a Cameca NanoSIMS 50L ion microprobe. To this end, Cr isotopic images (50, 52, 53, 54) of very large numbers of grains in the residue were collected. Because O isotopic compositions would be extremely helpful with determining the source of the grains, O isotopic images were collected for many of the same areas. In a preliminary study of some 200 Cr isotope images, we identified several $^{54}$Cr-enriched areas (Qin et al., 2009). However, implantation of $^{16}$O from the primary beam during the measurements precluded accurate determination of O-isotopic compositions. Thus, in subsequent work we acquired O isotopic images (spatial resolution ~150 nm) for an additional ~100 areas and reanalyzed most for Cr isotopic composition (Nittler et al., 2010). We note that Dauphas et al. (2010) recently reported very similar results.

## 2. Samples and Methods

The original Orgueil residue was prepared using CsF/HCl dissolution (Alexander et al., 2007; Cody et al., 2002; Cody and Alexander, 2005), followed by removal of carbonaceous material by O-plasma ashing. The remaining material (mostly small grains of chromite and spinel) was dispersed as a liquid suspension onto a high purity gold substrate. Scanning electron microscope imaging showed that the minerals in the sample mount were mostly sub-micron Cr oxides, primary chromite, and a small amount of SiC. The Cr and O isotopic compositions of the grains were determined with a Cameca NanoSIMS 50L ion microprobe in imaging mode, in analytical sessions using $O^-$ and $Cs^+$ primary ion sources, respectively. The density of the grains on the mount was very high, allowing for 100s to 1000s of grains to be analyzed in each image, but with the drawback that the isotopic compositions of single grains are in many cases strongly influenced by dilution by signals from surrounding materials.

*2.1. Cr isotope measurements*

A 300-700 nm primary $O^-$ ion beam of ~ 5-10 pA intensity was rastered over $25 \times 25$ μm$^2$ or $20 \times 20$ μm$^2$ areas that were pre-sputtered with a stronger beam for about 10 minutes to remove surface contamination. Positive secondary ions of $^{28}$Si, $^{48}$Ti, $^{50}$Cr + $^{50}$Ti, $^{52}$Cr, $^{53}$Cr, $^{54}$Cr + $^{54}$Fe and $^{56}$Fe were monitored simultaneously on 7 electron multipliers; Ti and Fe were monitored to correct for isobaric interferences from $^{50}$Ti and $^{54}$Fe on $^{50}$Cr and $^{54}$Cr, respectively. $^{28}$Si was monitored to aid in the distinction of SiCs or silicates from the major Cr-bearing minerals in the residue. For each area, 20-40 sequential 256 × 256 or 128 × 128 pixel images were obtained over a period of 40 minutes. Isotope ratio images were generated and isotopic ratios for individual grains were determined through image processing with the L'image software package (L. R. Nittler, Carnegie Inst.). The Cr isotope ratios were normalized to the average composition in each image.

$^{54}$Cr and $^{50}$Cr were corrected for interference from Fe and Ti, respectively, with the assumption of normal isotopic compositions for these elements. In our earliest measurements, $^{48}$Ti was not monitored. Consequently, the interference on $^{50}$Cr from $^{50}$Ti could not be corrected for in these measurements (see footnotes of Table 1). Because the measured isotopic ratios drifted with time and varied between images due to changes in instrumental mass fractionation and electron multiplier detection efficiency, a correction was determined internally for each image. Each image was divided into numerous small regions of interest (ROIs), for which isotopic ratios were determined and the raw measured ($^{50}$Cr+$^{50}$Ti)/$^{52}$Cr ratios were plotted against $^{48}$Ti/$^{52}$Cr ratios for all the ROIs in an image (Fig.1a). With the assumption that instrumental fractionation is the same for all the grains in the same image, the slope of a regression line to the data yields the actual $^{50}$Ti/$^{48}$Ti in the image, and the intercept with $^{48}$Ti/$^{52}$Cr=0 yields the average $^{50}$Cr/$^{52}$Cr in the image (Fig.1a). The fact that the raw data on ($^{50}$Cr+$^{50}$Ti)/$^{52}$Cr were always reasonably well correlated with $^{48}$Ti/$^{52}$Cr ratios, indicates that the method works as intended. The interference on $^{54}$Cr from $^{54}$Fe was corrected in a similar fashion as for Ti (Fig.1b). In this image, after interference-correction, most ROIs show δ ($^{54}$Cr/$^{52}$Cr) equal to zero within error, with the exception of one region with a positive anomaly.

The spatial resolution of our analyses is limited by the size of the primary O$^-$ beam. The typical analytical uncertainty (1σ) for individual (~500 nm-sized) regions on δ$^{50}$Cr, δ$^{53}$Cr and δ$^{54}$Cr is 15 ‰, 10 ‰ and 25 ‰, respectively.

*2.2. Oxygen isotope measurements*

Oxygen isotopic compositions were determined in an area of the sample mount that was not previously sputtered for Cr isotope analyses. This is because the O$^-$ beam used to measure Cr isotopes implants a large amount of $^{16}$O into the grains and this $^{16}$O-rich layer, which compromises the O isotope analyses, is very difficult to remove.

Oxygen isotopes were measured with a ~100 nm primary Cs$^+$ beam rastered on areas of 20 × 20 μm$^2$ and simultaneous collection of negative ions $^{16,17,18}$O$^-$, $^{28}$Si$^-$, $^{27}$Al$^{16}$O, and $^{52}$Cr$^{16}$O$^-$ on electron multipliers. 20 sequential 256 × 256 pixel isotope images were collected for each area. Counting times resulted in typical uncertainties of 2-10 % for O isotope ratios for sub-micron oxide grains. The O isotope ratios for individual grains were also normalized to the average composition in a given image. Many of the same areas that had been scanned for O isotopes were subsequently mapped for Cr isotopes with the methodology described above.

**3. Results**

*3.1. Cr isotope imaging results*

We identified a total of ten regions with $^{54}$Cr/$^{52}$Cr ratios that deviated from the mean by 3-20 σ, with δ$^{54}$Cr values ranging from ~100 ‰ to 1500 ‰ (Figs. 2-4, Table 1).

The $^{54}$Cr-rich areas show no resolvable anomalies in either $\delta^{50}$Cr or $\delta^{53}$Cr (Fig. 4), except for grain 7_5 that shows a 70 ‰ $^{53}$Cr deficit.

These $\delta^{54}$Cr anomalies are all statistically significant and cannot be due to an improper Fe correction caused by the presence of a $^{54}$Fe anomaly in the grains (Appendix 1, Fig. A1). The Cr-isotopic measurements of the $^{54}$Cr-rich grains are severely diluted by signals from surrounding material on the sample mount due to the relatively large size of the NanoSIMS O$^-$ primary ion beam (>500 nm in this study) compared to the size of the grains and the high grain density in the sample mount. Thus, these isotope anomalies are only lower limits, and many anomalous grains were almost certainly missed in our search, precluding an accurate determination of the abundance of $^{54}$Cr-rich grains in the Orgueil residue. The true abundance of $^{54}$Cr-rich grains in the Orgueil residue is certainly much larger than implied by our results (< 0.002 area %).

*3.2. Oxygen isotope imaging results*

Oxygen isotope imaging of ~ 100 areas, conducted prior to mapping them for Cr isotopic compositions, revealed 161 presolar oxide grains (Fig. 5). Most grains are < 400 nm in size and appear to be Al-rich. The grains span a similar range of O isotope compositions as previously observed for presolar oxides (Nittler et al., 2008) and silicates (Nguyen et al., 2007), and the majority of them were probably derived from asymptotic giant branch (AGB) stars. However, three grains with extreme $^{17}$O enrichments and two grains with extreme $^{18}$O enrichments may have formed in novae and supernovae, respectively (Nittler et al., 2008).

Three of the $^{54}$Cr-rich grains were identified in areas previously analyzed for O isotope compositions, including two of the most extreme grains, 7_10 and 8_1 (Figs. 2 and 3). Comparison of $^{52}Cr^+$ and $^{52}Cr^{16}O^-$ images collected during Cr and O isotope analytical sessions, respectively, allowed for the extraction of some O isotope information for these $^{54}$Cr-rich grains (Fig. 2). The most $^{54}$Cr-rich grain, 7_10 (Fig. 2), had too few O ion counts to be useful, and the other two grains have O isotope compositions that are within error of normal (Table 1), ruling out large $^{17}$O- or $^{18}$O-enrichments. However, as will be discussed later, enrichments in $^{16}$O for all three grains cannot be ruled out because $^{17}$O and $^{18}$O depletions are more easily masked by contributions from surrounding isotopically normal grains.

*3.3. SEM and Auger results*

Scanning electron microscope (SEM) examination of the areas analyzed by NanoSIMS revealed that in most cases the anomalous regions are associated with multiple grains, making it impossible to determine which carries the Cr anomaly. However, in a few cases the $^{54}$Cr anomaly can be clearly identified with a single grain (Fig. 2), although they are still not well separated from other grains on the mount. The anomalous grains were too small (< 200 nm) to be reliably analyzed with the SEM. However, data from both the NanoSIMS images and subsequent Auger analyses indicate that the grains contain Cr, O, and in some cases other elements including Al, Fe and/or Ti (Figs. 2d and 3d). They thus appear to be Cr-bearing oxides, perhaps spinels. Although such phases should be resistant to hot HCl leaching, such tiny grains could be at least partially dissolved if they contain defects, or if their surfaces were radiation damaged (Dauphas et al., 2010).

*3.4. NanoSIMS image simulations*

As previously discussed, the Cr-isotopic measurements of the $^{54}$Cr-rich grains are severely diluted by signals from surrounding material on the sample mount. To quantitatively estimate the magnitude of this effect, we used higher-resolution images acquired either with the NanoSIMS Cs$^+$ beam or a SEM to generate simulated Cr-isotopic images for three of the anomalous grains. We varied the assumed $^{54}$Cr/$^{52}$Cr ratios in the simulations in order to match the actual measurements.

Grains 7_10 and 8_1 were identified in areas that were first analyzed for O isotopes with the NanoSIMS Cs$^+$ beam (spatial resolution ~150 nm). We used the CrO$^-$ secondary ion images acquired during these measurements as the basis for simulated Cr images, since CrO$^-$ ion yields from oxide grains are correlated with Cr contents (see below). The CrO$^-$ and Cr$^+$ images for the area 7_10 are shown in Figs. 6a and 6b; because the measurements were made at different times, there is a slight offset in the actual analyzed areas. The anomalous grain in this image is identifiable as a ~100 nm grain next to a larger grain (arrow in Figs. 2 and 6b). A synthetic, high-resolution $^{52}$Cr image was first generated by scaling the CrO$^-$ image to the observed average count rate in the Cr$^+$ image for the same area. A $^{54}$Cr image was next generated by scaling the $^{52}$Cr image by the terrestrial $^{54}$Cr/$^{52}$Cr ratio (0.0282), with the exception of the pixels corresponding to the anomalous grain. The number of $^{54}$Cr counts in these pixels was varied to correspond to a range of true $^{54}$Cr/$^{52}$Cr ratios. Next, for each assumed $^{54}$Cr/$^{52}$Cr ratio, the high-resolution Cr isotope images were convolved with a 2-dimensional Gaussian to match the spatial resolution of the observed Cr$^+$ images (FWHM=650 nm for 7_10). A resulting simulated $^{52}$Cr image for 7_10 is shown in Fig. 6c. Finally, Poisson noise was added to the simulated images, and then they were processed with the same software used to analyze the real data. In particular, $^{54}$Cr/$^{52}$Cr ratios were determined for regions of interest corresponding to the same size as those used to define the grains in the original Cr-isotopic images. Comparison of these values to the real measurements then allowed us to determine the spread in possible true compositions consistent with the data. For grain 7_10, we obtain the best match to the observations if the 100 nm grain has a true $\delta^{54}$Cr/$^{52}$Cr ratio =44,000±6,000 ‰ (Fig. 6e), corresponding to the ±1σ error for the original measurement (Table 1).

This procedure is based on the assumption that the CrO$^-$ signals acquired with a Cs$^+$ beam can be directly related to the Cr$^+$ signals acquired with an O$^-$ beam. To test this assumption, we compared the count rates for individual grains in the 7_10 $^{52}$Cr$^+$ image (Fig. 6b) to those in the simulated image (Fig. 6c). We found that for 50 grains spanning a wide range of count rates, the simulated and observed count rates were well correlated with each other, with a standard deviation of ~35%. That is, the true $^{52}$Cr count rate can deviate from the rate predicted by Gaussian-smoothing of the CrO$^-$ image by ± 35% (1σ), providing an additional source of systematic uncertainty. Thus, a ~3σ lower limit on the inferred $^{54}$Cr/$^{52}$Cr ratio for grain 7_10 can be obtained by assuming that the true $^{52}$Cr count rate for the anomalous grain is ~2.1 times the value derived from the CrO$^-$ image. Simulated images which have this $^{52}$Cr rate in the anomalous grain and which match the observed $^{54}$Cr/$^{52}$Cr ratio image have a true $\delta^{54}$Cr/$^{52}$Cr value for the grain of ~20,000 ‰, which we thus take as a robust lower limit on the true composition of this grain.

A similar procedure was applied to the $^{54}$Cr-rich grains identified in runs 8_1 and 8_3. For grain 8_1, the simulations are less certain due to the presence of three unresolved grains in the region of the $^{54}$Cr enrichment (Fig. 3). For 8_3, no high-resolution CrO$^-$ image was obtained, so we estimated the degree of isotopic dilution based on an SEM image. A binary 256x256 pixel mask image was first generated by re-sampling an SEM image and setting pixels of overlapping grains to a value of 1. This image is then used as a template to generate simulated isotopic images. For both grains, achieving a reasonable match to the observed $^{54}$Cr/$^{52}$Cr ratios requires true $\delta^{54}$Cr >10,000 ‰.

Extrapolation of the observed $^{53}$Cr depletion in grain 7_5 to $\delta^{53}$Cr = −1000 ‰ (corresponding to $^{53}$Cr/$^{52}$Cr=0) constrains the maximum true $\delta^{54}$Cr/$^{52}$Cr value for this grain to be <4,300 ‰ (2σ). Thus, there appears to be at least two types of anomalous grains, one with extreme $^{54}$Cr enrichments ($\delta^{54}$Cr >10,000 ‰) and one with less extreme $^{54}$Cr/$^{52}$Cr ratios, but $^{53}$Cr depletions.

*3.5. O isotopes in $^{54}$Cr-rich grains*

We obtained O isotopic data for three of the $^{54}$Cr-enriched grains, but low total counts preclude a meaningful result for Grain 7_10 (Table 1). The other two grains, 7_5 and 8_1, are within measurement errors of terrestrial O-isotopic compositions. However, isotopic data for both of these grains are compromised by dilution from adjacent material. As discussed in the previous section (Fig. 3), the $^{54}$Cr enrichment in run 8_1 corresponds to one of three unresolvable grains, and the reported O-isotopic composition ($\delta^{17}$O = −70±130 ‰, $\delta^{18}$O = 30±60 ‰) corresponds to a sum over image pixels covering all three grains. The minimum amount of O-isotopic dilution affecting the measurement can be estimated from the relative cross-sectional areas of the three grains under the SIMS beam.

If the largest grain, making up ~1/2 of the area (Fig. 6), is the anomalous one, then its true composition could be as $^{17}$O and $^{18}$O depleted as $\delta^{17}$O~ −700 ‰, $\delta^{18}$O~ −200 ‰ and still be compatible with the measured data at 2σ. If the anomaly is carried by one of the smaller ones, with ~1/4 of the area, then the true composition could be $\delta^{17}$O~ −1000 ‰, $\delta^{18}$O~ −360 ‰. The true effect of isotopic dilution will be worse than this, however, due to the tails on the ion beam (Nguyen et al., 2007).

## 4. Discussion

### 4.1. Sources of $^{54}$Cr-rich grains

Energetic particle induced spallation reactions on target elements like Fe and Ni can produce large excesses in $^{54}$Cr (and to a lesser extent $^{53}$Cr) if the material is also low in Cr (Qin et al., 2010a; Qin et al., 2010b). Orgueil has a low exposure age and a moderate Fe/Cr ratio. Consequently, any irradiation would have had to take place in the proto-planetary disk, in which case the best target grains would be those composed of Fe-Ni metal. Spallation of Fe and Ni will contribute to all four Cr isotopes in the proportions 0.2 : 1 : 1 : 1 for $^{50}$Cr, $^{52}$Cr, $^{53}$Cr (>80% from $^{53}$Mn decay) and $^{54}$Cr, respectively (Birck and Allègre, 1985; Qin et al., 2010a; Shima and Honda, 1966). Pure spallogenic Cr has an isotopic composition of $\delta^{54}$Cr=34,430‰, $\delta^{53}$Cr=7,820‰ and $\delta^{50}$Cr=2,860‰. Some of these values are not compatible ($\delta^{53}$Cr) or just barely compatible ($\delta^{50}$Cr) with extrapolation of our simulated values for grain 7_10 (Fig. 7). An additional step would be needed to extract the spallogenic Cr from a Fe-Ni-rich target and incorporate it in to the grains analyzed here without dilution from isotopically normal Cr. If this could have been done shortly after irradiation (i.e. before significant $^{53}$Mn decay), it would reduce the pure spallation $\delta^{53}$Cr value to ≤760 ‰, bringing the pure spallation composition closer to our extrapolated composition for grain 7_10. Such an alteration of the target grains would almost certainly have to occur on the parent bodies.

Very little Cr is actually created by spallation, so that to generate large anomalies requires very low initial Cr/Fe ratio in the target grains. The heavily irradiated iron meteorite Carbo has phases with very low Cr/Fe ratios of $10^{-6}$ to $10^{-5}$. Nevertheless, the maximum $\delta^{54}$Cr value in Carbo is ≤80 ‰ (Qin et al., 2010b), more than an order of magnitude lower than the more anomalous grains we report here, and almost three orders of magnitude less than our best estimate for the true composition of grain 7_10. Metal in chondrites is always Cr-bearing and any alteration in a parent body would almost certainly result in additional mixing with isotopically normal Cr. Hence, the spallation Cr would be diluted and one would expect $^{54}$Cr enrichments that were much less than estimated for the three anomalous grains and probably less than measured (Table 1). Thus, spallation would require several extremely unlikely circumstances to explain the grains we have found.

Slow neutron-capture nucleosynthesis in asymptotic giant branch (AGB) stars is expected to produce a small amount of neutron-rich isotopes like $^{54}$Cr. However, model calculations predict a maximum $^{54}$Cr excess in AGB stars of ~500 ‰ (Zinner et al., 2005), much lower than even the actual measured values of our most extreme grains. Thus, an AGB star origin for the grains can be ruled out.

The most likely explanation for the $^{54}$Cr-rich grains is that they formed in a supernova (SN). Chromium-54 and other neutron-rich nuclides, such as $^{50}$Ti, $^{48}$Ca, and $^{60}$Fe, are believed to form largely in neutron-rich environments in both Type Ia and Type

II SNe (Hartmann et al., 1985; Meyer et al., 1996). There are large uncertainties in models of both types of SNe, but we favor a Type II origin for the anomalous grains for a number of reasons. (1) The maximum $^{54}$Cr/$^{52}$Cr ratio produced in typical Type Ia SNe is ~ 4 × solar (Iwamoto et al., 1999), based on overall production of Fe-peak elements and Galactic Chemical Evolution considerations. This value is insufficient to explain the estimated $^{54}$Cr/$^{52}$Cr ratio for the most $^{54}$Cr-rich grains (> 11 × solar). (2) Much larger $^{54}$Cr abundances ($^{54}$Cr/$^{52}$Cr >40 times the solar ratio) are predicted by certain models of high-density Type Ia SNe (Woosley, 1997). However, such SNe are expected to be rare (~2% of Type Ia SNe) and, thus, are unlikely to be significant contributors to dust in the Galaxy. Moreover, these models generally predict very large abundances of other neutron-rich nuclei as well, especially $^{48}$Ca. Moynier et al. (2010) recently reported Ca-isotopic measurements of Orgueil leachates and found no evidence for $^{48}$Ca enrichments correlated with the known $^{54}$Cr anomalies. Assuming that the grains we have identified are indeed the source of the $^{54}$Cr enrichments in leachates, this is further evidence against a Type Ia SN source. (3) Type Ia SNe in general produce very little oxygen overall and, in Type Ia models that produce extreme $^{54}$Cr overabundances (e.g., Woosley 1997), the $^{54}$Cr is synthesized in a region without O, making it difficult to conceive how oxide grains like those reported here could form. In contrast, in Type II SNe very high $^{54}$Cr/$^{52}$Cr ratios, accompanied by smaller $^{53}$Cr excesses and $^{50}$Cr depletions, are predicted for the $^{16}$O-rich O/Ne and O/C zones (Woosley and Heger, 2007) providing a direct route to oxide formation.

The predicted Cr isotopic ratios for different zones of a 20 $M_\odot$ Type II SN (Woosley and Heger, 2007) are shown in Fig. 7. The precise ratios predicted for a given

SN zone will depend both on the mass of the star and the particular model employed (and its associated uncertainties), but the nuclear processes involved are well understood and the general isotopic features shown by the Figure should be robust. The values for the O/Ne and O/C zones are in good agreement with our estimated true composition of grain 7_10 (Table 1). Extrapolation of the $^{53}$Cr/$^{52}$Cr and $^{54}$Cr/$^{52}$Cr ratios for grain 7_5 is consistent with formation in the O/Si zone of the same SN model (dashed curves). If these grains are indeed from the $^{16}$O-rich O/Ne, O/C and/or O/Si zones of Type II SNe, very low $^{17}$O/$^{16}$O and $^{18}$O/$^{16}$O ratios ($\delta^{17}$O ~$\delta^{18}$O ~ -1000) are expected (Fig. 5). As discussed earlier, although we did not observe any O isotope anomalies associated with the $^{54}$Cr-rich grains, we cannot rule out that they are $^{16}$O-rich because the $^{16}$O-rich signature can be easily masked both by low counting statistics and by surrounding isotopically normal grains. We thus believe that the measured O-isotopic compositions of the grains are consistent with the $^{16}$O-rich compositions expected if the $^{54}$Cr anomalies reflect an origin in the $^{16}$O-rich zones of a Type II supernova. Moreover, if these grains are from SNe, other neutron-rich isotopes may also have been incorporated into the protoplanetary disk along with $^{54}$Cr. As mentioned earlier, the low Ti abundance and isobaric interference from $^{50}$Cr, make it very difficult to search for possible correlated anomalies in $^{50}$Ti in the $^{54}$Cr-rich oxide grains. However, the low Ti abundances in the $^{54}$Cr-rich grains exclude them as the major carrier of $^{50}$Ti anomalies found in meteorites. There are rare Ti-rich grains in our Orgueil residue. A Ti isotopic study of these Ti-rich grains may provide a test of whether the $^{54}$Cr and the $^{50}$Ti anomalies formed in the same SN sources.

Although most known presolar oxide and silicate grains are believed to have formed around AGB stars, a sub-population of grains, mostly $^{18}$O-rich, likely formed in Type II supernovae (Bland et al., 2007; Busemann et al., 2009; Gyngard et al., 2010; Nguyen et al., 2007; Nguyen et al., 2010; Nittler et al., 2008; Vollmer et al., 2009) (grey symbols in Fig. 5). A major fraction of these SN grains appear to lie on a single line on the O 3-isotope plot (Fig. 5). This line has been interpreted as a mixing line between different supernova zones, perhaps reflecting a jet of $^{16}$O-rich material from inner regions passing through and mixing with a pre-existing mixture of outer zones (Nittler et al., 2008). Hence, the population of known O-rich supernova grains is dominated by grains that formed under a rather narrow range of mixing conditions. Grains dominated by inner-zone material, as we have inferred for grain 7_10 (Fig. 5), should also plot on this mixing line. Astronomical observations, as well as the isotopic compositions of presolar SiC and graphite grains, show that the mixing in SN ejecta is highly variable within and between remnants (Hoppe et al., 2000; Hwang et al., 2004; Park et al., 2004; Travaglio et al., 1999). Given the extreme diversity in SN compositions, a scenario in which multiple SN ended up producing grains that primarily lie along a single O isotope mixing line as observed here is highly unlikely. Nittler et al. (2008) thus argued that it is much more plausible that these oxide grains formed in a single SN, and we propose that the $^{54}$Cr-rich nanoparticles likely formed in the same stellar explosion. However, this conclusion rests on the assumption that the grains have large $^{16}$O enrichments and lie on the same mixing line on the O 3-isotope plot, which clearly needs to be confirmed with future measurements of similar grains. Moreover, additional isotopic measurements of other elements in both the $^{54}$Cr-rich grains and the $^{17,18}$O-rich grains are highly desirable to test the validity of this model.

We note that there is no strong evidence for a dominant single SN source for other known types of presolar SN grains, like SiC, graphite and $Si_3N_4$, indicating that they most likely formed in multiple SN sources. This suggests that the SN that contributed the oxide and silicate grains along the O isotope mixing line did not produce significant amounts of reduced phases. Also, although we argue that the SN oxide and silicate population is dominated by grains from a single SN, some of the grains clearly do not lie along the O-isotope mixing line and thus likely formed in different SNe, perhaps the same ones which produced the SN SiC, graphite and $Si_3N_4$ grains.

*4.2. Implications for the $^{54}$Cr variations in the Solar System*

A previous study suggested that the distribution of $^{54}$Cr-carriers is controlled by volatility, and that the carrier phase preferentially resides in the matrix (Shukolyukov and Lugmair, 2006). However, we have found that there is no clear correlation between matrix abundance and $^{54}$Cr-excess in bulk chondrites (Qin et al., 2010a). To further test this, individual chondrules from the relatively primitive CO chondrite Ornans, and CR chondrite EET 92042 were studied. The results are shown in Table 2. Ornans chondrules show heterogeneous $\varepsilon^{54}$Cr ranging from 0.54±0.21 to 1.11±0.1. The bulk CO chondrites have $\varepsilon^{54}$Cr values from 0.5 to 0.8 (Qin et al., 2010a; Trinquier et al., 2007). Thus, at least one chondrule has an excess compared to the bulk CO meteorites. All three EET 92042 chondrules have the same $\varepsilon^{54}$Cr value ~ 1.30 (Table 2), which is identical to bulk CR chondrite data. These results argue against the $^{54}$Cr carrier phase preferentially residing in the matrix. Thus, the bulk chondrite variations are probably not controlled by the volatility of the sample or the abundance of matrix, but must reflect heterogeneity in the solar nebula that all components in a particular chondrite group sampled.

Despite their small sizes, the extreme $^{54}$Cr enrichments of the grains observed here indicate that they could be a major contributor to $^{54}$Cr variations in bulk chondrites. Most of the observed variation is in $^{54}$Cr, but one grain also shows a moderate $^{53}$Cr depletion that suggests that at least part of the anti-correlation between $^{53}$Cr and $^{54}$Cr seen in leaching experiments (Podosek et al., 1997; Qin et al., 2010a; Rotaru et al., 1992) and in the bulk Orgueil residue (Qin et al., 2010a) can be attributed to similar grains. Note that mass balance calculations indicate that some $^{54}$Cr-rich phases were dissolved during preparation of the residues, and therefore, there are likely to be multiple carrier phases, perhaps including SN silicates (Qin et al., 2010a) that probably formed in the same SN as the oxide grains.

Assuming that the $^{54}$Cr-rich grains are indeed of SN origin, the bulk Cr isotopic variations between different meteorite classes point to significant spatial and/or temporal heterogeneity in the abundance of these grains in the solar protoplanetary disk. The relative abundances of the $^{18}$O- or $^{16}$O-rich SN oxide and silicate grains among different types of primitive material are not very well known, but extant data for silicates are summarized in Table 3. Presolar silicates are highly susceptible to destruction by parent-body processing and thus are only found in high abundances in the most primitive (type 3.0) meteorites (Floss and Stadermann, 2009; Nguyen et al., 2007). Supernova ($^{18}$O-rich or $^{16}$O-rich) grains appear to be present at a constant relative abundance of ~13% across the four best-studied primitive chondrites (Table 3), though the uncertainties arising from poor statistics allow for some heterogeneity. Unfortunately, bulk Cr isotopic data are not available for these meteorites. Interestingly, for interplanetary dust particles and Antarctic meteorites, the relative abundance of SN silicate grains appears to be ~3 times higher than the chondrites. Since these classes of extraterrestrial materials sample different parent bodies, possibly comets (Duprat et al., 2010; Messenger et al., 2005), than the larger meteorites, this also points to a heterogeneous distribution of SN-derived O-rich grains in the disk. Statistics for presolar grains in the comet Wild-2 samples returned by the Stardust mission are too poor to provide meaningful insight into this question.

Most presolar silicate, oxide, SiC and graphite grains had multiple stellar sources, so it seems likely that these presolar grains were well-mixed components of the protosolar molecular cloud. In general, the relative abundances of different types of presolar grains in the most primitive meteorites are fairly constant (Huss and Lewis, 1995; Huss et al., 2003), despite grain sizes that range from ~2–5 nm to several μm. What variations there are can largely be ascribed to parent body processes and uncertainties in the measurements (Huss et al., 2003). The $^{54}$Cr-rich grains we found are only slightly smaller than typical presolar silicates and oxides and, hence, physical processes in the disk are unlikely to have fractionated the $^{54}$Cr-rich grains to produce the bulk meteorite Cr isotopic variations without also affecting the other types of presolar grains. Moreover, chemical processes in the disk are not likely to have fractionated the $^{54}$Cr-rich grains because the grains are chemically relatively stable.

Thus, the $^{54}$Cr-rich and most other SN oxide grains seem to be decoupled from most other known presolar grains. If there are no physical or chemical means of decoupling the $^{54}$Cr-rich grains, this raises the intriguing possibility that they were introduced into the Solar System at a later time than other presolar grains, probably during formation of the Solar System. Numerous authors have appealed to either supernova triggering of Solar System formation (Boss, 2004; Cameron and Truran, 1977; Vanhala and Boss, 2000; Vanhala and Boss, 2002), or injection of supernova material into the already formed protoplanetary disk (Clayton, 1977; Ouellette et al., 2007) to explain some or all of the short-lived radionuclides that were present in the early Solar System. Both mechanisms would produce spatial and temporal heterogeneities in the disk.

However, recent hydrodynamic models show that in the early stages of Solar System formation, large spatial heterogeneities would have been erased to a level of about 2-10% relatively quickly, on timescales of ~1000 yrs in a SN triggering collapse of the protosolar cloud (Boss, 2007; Boss, 2008; Boss, 2009). There are large uncertainties in the abundance and isotopic composition of the $^{54}$Cr-rich carrier in the meteorites, and in the initial Cr isotopic composition of the protosolar system prior to the SN injection. However, we found that to determine what variability in the distribution of the $^{54}$Cr-rich phase(s) can account for the variations in $^{54}$Cr abundances among difference meteorite classes, only the initial Cr isotopic composition in the protosolar system is needed. The first one or two weak acid steps of bulk chondrite leaching experiments on relatively primitive meteorites reveal -1 ~ -0.5 ‰ $\delta^{54}$Cr deficits (Podosek et al., 1997; Trinquier et al., 2007). If this is the Cr isotopic composition before the SN injection, mass balance calculations show that a 20 % to 50 % variation in the distribution of the $^{54}$Cr-rich carrier is required to account for the total variations of > 0.2 ‰ observed between different meteorites classes, larger than the <10% level predicted by the hydrodynamic models.

The initial $\delta^{54}$Cr value in the solar nebula prior to SN injection could well be larger, if we assume that some of the acid leachable $^{54}$Cr-rich components in step-wise leaching experiments are not SN oxides (these grains are presumably homogeneously distributed between different meteorite classes). The larger the initial $\delta^{54}$Cr value of the solar nebula is, the larger the degree of the heterogeneity in the distribution of the $^{54}$Cr-rich SN carrier must be. Moreover, the time difference between the onset of collapse of the Solar System and the formation of different planetesimals is probably of the order of a few million years, by which time even the small residual heterogeneities allowed by the hydrodynamic models (Boss 2008) are likely to have been erased. Thus, the most plausible explanation is that there was an injection of SN material into an already forming disk. Recent model calculations suggest that 0.1 to 10 μm supernova dust grains would be the most efficiently injected into a nearby protoplanetary disk (Ouellette et al., 2010), whereas gas from the SN will flow around the disk and not be injected. The size of the observed $^{54}$Cr-rich grains is within this range. Recent simulations indicate that formation of planetesimals from millimeter-sized particles could have been a very turbulent and rapid process (Johansen et al., 2007; Johansen et al., 2009), thereby preserving a record of any local heterogeneities in the disk.

HEDs, pallasites, ureilites and iron meteorites have the lowest bulk $\delta^{54}$Cr values amongst all meteorite types, roughly -0.07 ‰, a smaller deficit (-0.03 ‰) has been reported for angrites (Qin et al., 2010a; Qin et al., 2010c; Trinquier et al., 2007). Carbonaceous chondrites show varying positive anomalies in $\delta^{54}$Cr from 0.04 ‰ to 0.16 ‰, while O- and E-chondrites have intermediate values of -0.04 ‰ and 0 ‰, respectively (Qin et al., 2010a; Shukolyukov and Lugmair, 2006; Trinquier et al., 2007). Martian meteorites show a very small deficit of about -0.02 ‰ (Qin et al., 2010a; Trinquier et al., 2007). The small difference between Earth and Mars argues against a large radial heterogeneity of the $^{54}$Cr carrier in the inner Solar System. The deficit in $^{54}$Cr-rich components in most differentiated meteorites may indicate that their parent bodies formed either prior to injection of the SN grains, or in regions of the protoplanetary disk with higher surface densities that led to greater dilution of the $^{54}$Cr-rich components. Short-lived chronology based on Hf-W ages of iron meteorites (Markowski et al., 2006; Qin et al., 2008), and Al-Mg and Mn-Cr ages of eucrites (Bizzarro et al., 2005; Lugmair and Shukolyukov, 1998; Trinquier et al., 2008) is consistent with the parent bodies of these differentiated meteorites accreting within the first 1 or 2 Myr after formation of the Solar System, likely predating the formation of chondrites judging from Al-Mg and Pb-Pb ages of chondrules (Amelin et al., 2002).

*4.3. Collateral consequences of SN injection*

Injection of grains from a nearby SN into the protoplanetary disk would undoubtedly have collateral consequences beyond Cr isotopes and the distribution of presolar grains. A key feature of the model is the selective sampling of the SN ejecta implied by the presolar grain data (e.g., the particular mixing of certain SN regions but not others). This is in strong contrast to previous models (Gounelle and Meibom, 2007; Nichols et al., 1999), which assumed bulk SN yields for calculating collateral isotope effects of late SN injection. In the present model, the isotopic composition and abundance of any element injected from the SN will depend on the precise mixtures of zones included in the mix, as well as the abundances, types and size distributions of dust grains formed and the relative survival of different dust types upon injection into the Solar System. Since dust formation in supernovae is poorly understood, this means that any quantitative predictions of collateral effects will be highly uncertain at this point and a detailed discussion of such effects is beyond the scope of this paper. However, we do briefly consider the cases of Ti and Ni isotopes, since these have been reported to correlate with $^{54}$Cr anomalies in bulk meteorites (Regelous et al., 2008; Trinquier et al., 2009), as well as the important extinct radionuclide $^{60}$Fe.

The *s*-process occurring in the O/Ne and O/C zones of a Type II SN produces excesses in all Ti isotopes, relative to $^{48}$Ti and, as we have seen earlier (Fig. 7) depletes $^{50}$Cr and enriches $^{53}$Cr and $^{54}$Cr. Here we explore the consequences of mixing variable amounts of the O/Ne and O/C zones of the 20 $M_\odot$ SN model of Woosley and Heger (2007) with a bulk solar composition. We make the simplifying assumptions that there is no element fractionation in the bulk dust condensed in these zones and that different types of dust survive injection into the disk in the same relative proportions. We calculate $\varepsilon^{54}$Cr, $\varepsilon^{46}$Ti and $\varepsilon^{50}$Ti values from the mixed compositions, applying the same exponential mass fractionation laws used in the bulk meteorite studies (Qin et al., 2010a; Trinquier et al., 2009). This allows direct comparison with the bulk data, which have been analyzed with the assumption of normal $^{50}$Cr/$^{52}$Cr and $^{49}$Ti/$^{47}$Ti ratios. Results are shown in Fig. 8, for a range of mixing fractions of the O/Ne zone of 0 to $8\times10^{-5}$ by mass; very similar results are obtained for the O/C zone. The model predicts correlations between $\varepsilon^{54}$Cr, $\varepsilon^{46}$Ti and $\varepsilon^{50}$Ti, in good agreement with the observations. The slope of the $\varepsilon^{50}$Ti versus $\varepsilon^{54}$Cr correlation line is shallower (~0.5, Fig. 8a) than that reported by Trinquier et al (2009) for solar system bodies (1.4). However, the latter is somewhat uncertain, due to the need to correct the bulk carbonaceous chondrite data for the presence of calcium-aluminum rich inclusions. Moreover, the correct slope could easily be reproduced if there was a fractionation between Ti and Cr in the SN dust such that the Ti/Cr ratio of the dust was higher than the bulk SN zone gas. The predicted correlation between $\varepsilon^{46}$Ti and $\varepsilon^{50}$Ti is in much better agreement with the observed trend (Fig. 8b). Note that when ratioed to the most abundant isotope $^{48}$Ti, the mixtures are in fact strongly enriched in $^{49}$Ti and $^{50}$Ti with smaller anomalies in $^{46}$Ti and $^{47}$Ti; the relatively large $^{46}$Ti anomalies seen here (and by extension in the bulk meteorite data) are a consequence of correcting the data with the assumption of a normal $^{49}$Ti/$^{47}$Ti ratio. The rather remarkable agreement, considering the large uncertainties, between the predicted Cr-Ti correlations and the observed planetary data supports our contention that the bulk meteorite anomalies reflect variable amounts of material from the O/Ne and O/C zones of a Type II supernova.

Regelous et al (2008) reported a correlation between $^{62}$Ni/$^{58}$Ni ratios and $^{54}$Cr/$^{52}$Cr ratios for bulk meteorites, though the magnitude of the observed $^{62}$Ni effect is some five times smaller than that of $^{54}$Cr. The SN zones considered here are indeed enriched in $^{62}$Ni, due to the *s*-process, but interestingly are even more enriched in $^{61}$Ni (and strongly depleted in the normalization isotope $^{58}$Ni). For example, in the O/Ne zone of the 20 M$_\odot$ SN model of Woosley and Heger (2007), the predicted $^{61}$Ni/$^{58}$Ni= $2.6\times10^4$ times solar and $^{62}$Ni/$^{58}$Ni= $1.1\times10^4$ times solar, both significantly larger than the relative enrichment of $^{54}$Cr/$^{52}$Cr (Fig 7). This means that if, as assumed above for Ti, the Ni/Cr ratio is unfractionated in the SN dust, the injection model would predict significantly larger isotopic effects for Ni than for Cr. Moreover, since the $^{61}$Ni/$^{62}$Ni ratio is higher than solar in the SN zone, correcting the data with the assumption of normal $^{61}$Ni/$^{58}$Ni (as done by Regelous et al, 2008), would lead to apparent $^{62}$Ni *depletions*. Therefore, with the oversimplification of no element fractionation, our model does not appear to be consistent with the reported bulk Ni isotopic data. However, equilibrium condensation calculations based on a similar SN model to the one we consider (Fedkin et al., 2009) predict Ti and Cr to condense into refractory oxides, but Ni into metallic alloys, in the O/Ne and O/C zones. The different condensation temperatures and material properties of oxides versus metal thus provide a route to element fractionation in the injection model and we do not consider the Ni data to be a strong test of the proposed model.

The extinct radionuclide $^{60}$Fe is also produced in these zones. However, the amount of $^{60}$Fe predicted to accompany the $^{54}$Cr-rich material depends critically on the assumed initial $^{54}$Cr/$^{52}$Cr ratio of the solar system. Dauphas et al. (2010) assumed that $^{54}$Cr rich grains get mixed into a solar system with terrestrial $^{54}$Cr/$^{52}$Cr and found only a 10-20% perturbation on the assumed initial $^{60}$Fe/$^{56}$Fe ratio. In contrast, if the SN grains are mixed into a disk with an initial 0.5 to 1 ‰ depletion in $^{54}$Cr, as assumed above, a larger fraction of SN material must be incorporated leading to higher $^{60}$Fe/$^{56}$Fe ratios (>10$^{-6}$) than inferred for the early solar system (Mishra et al., 2010). Note that in the direct injection model the time between the SN explosion and the injection into the disk is very short (Ouellette, et al., 2007) so we assume no decay of $^{60}$Fe in the calculation. In the O/Ne and O/C zones, Fe is expected to condense primarily into silicates (Fedkin et al., 2009); if these are produced or survive injection less efficiently than Cr- or Ti-bearing oxides, a smaller injection of $^{60}$Fe would be predicted.

The case of short-lived $^{26}$Al is much more complicated, because it is produced both in the $^{54}$Cr-rich zones and the outer zones which dominate the compositions of the $^{18}$O-rich presolar SN grains. Moreover, the SN injection would also lead to non-mass-dependent variations in the stable Mg isotopes, complicating the interpretation of bulk Al-Mg data for meteorites. A detailed analysis of this issue will be the subject of future work.

These results underscore the difficulty of making quantitative predictions of collateral isotopic effects of the proposed model. Additional data for SN oxide and silicate grains should provide better constraints on the range of mixing parameters for different SN zones and on the types of grains produced in Type II SNe and thus put firmer constraints on possible consequences of the proposed model.

## 5. Conclusions:

NanoSIMS imaging of mineral separates from an acid-resistant residue of Orgueil revealed 10 regions with positive $\delta^{54}$Cr anomalies up to 1500 ‰. These regions are not associated with $\delta^{50}$Cr and $\delta^{53}$Cr anomalies, except one region where a small negative $\delta^{53}$Cr anomaly was detected. These $\delta^{54}$Cr anomalies are clearly resolvable with our analytical uncertainties, and cannot be caused by analytical artifacts. The anomalies are also very unlikely to reflect $^{54}$Fe-rich supernovae grains. Because of dilution from surrounding normal grains, the measured anomalies are lower limits, and we estimated that the actual $^{54}$Cr/$^{52}$Cr enrichments in the grains may be as high as 50 times solar. Such enhancements in $^{54}$Cr/$^{52}$Cr ratio cannot be produced by early irradiation of dust by energetic particles from the young Sun, or in Type Ia supernovae, but strongly favor an origin in Type II supernovae.

Oxygen isotope imaging of a subarea mapped for Cr isotopes revealed 161 presolar oxide grains. The grains span a similar range of O isotope compositions as previously observed for presolar oxides (Nittler et al., 2008) and silicates (Nguyen et al., 2007), and the majority of them were probably derived from asymptotic giant branch (AGB) stars. However, three grains with extreme $\delta^{17}O$ enrichments and two grains with extreme $\delta^{18}O$ enrichments may have formed in novae and supernovae, respectively (Nittler et al., 2008). For three $\delta^{54}Cr$ rich grains, O isotope data were extracted and did not show resolvable anomalies. However, negative $\delta^{17}O$ and $\delta^{18}O$ isotope anomalies are expected if they are indeed from the inner zones of Type II supernovae, and cannot be ruled out because negative anomalies are most vulnerable to dilution from surrounding minerals.

SEM imaging showed that the $\delta^{54}Cr$-rich regions are often associated with more than one grain. However in a few cases the anomalous grains were well separated from other grains - they are usually < 200 nm in size and smaller than the typical grain size in the sample mount. Data from both the NanoSIMS images and subsequent Auger analyses indicate that the grains contain Cr, O, and in some cases other elements including Al and/or Ti. They are likely to be Cr-bearing spinels.

The $^{54}Cr$-rich grains, along with other SN-derived oxide/silicates grains plot on a single SN mixing line. Given the extremely complicated mixing conditions in supernova ejecta and tremendous variations observed in the isotopic compositions of SN-derived SiC grains, the unusually limited distributions of O isotopic compositions of these grains are most plausibly explained by a single supernova injection event. These supernova grains seem decoupled from other types of presolar grains, suggesting that they were injected later into the solar nebula.

Despite their small sizes, the extreme $^{54}$Cr enrichments of the grains observed here indicate that they could be a major contributor to $^{54}$Cr variations in bulk chondrites. We suggest that the variability in bulk $^{54}$Cr/$^{52}$Cr between meteorite classes reflects heterogeneous distribution of the $^{54}$Cr carrier phases in the already-formed solar protoplanetary disk following a single late supernova injection event. Detailed consequences of this suggestion will be the focus of future work.

**Acknowledgements:** We thank Nicholas Moskovitz for discussions, M.-C. Liu for help with the instruments, and Stan Woosley and Alex Heger for providing supernova yields in digital form. Comments from Jamie Gilmour, Alex Shukolyukov and an anonymous reviewer are greatly appreciated. Qin acknowledges support in the form of a postdoctoral fellowship from Carnegie Institution of Washington. This work was supported by NASA Cosmochemistry Grants NNX08AH65G and NNX07AJ71G.

Alexander C. M. O'D. and Carlson R. W. (2007) How homogeneous was the early Solar System? The chromium story. *Workshop on Chronology of Meteorites and the Ealry Solar System*, #4046 (abst.).

Alexander C. M. O'D., Fogel M., Yabuta H. and Cody G. D. (2007) The origin and evolution of chondrites recorded in the elemental and isotopic compositions of their macromolecular organic matter. *Geochim. Cosmochim. Acta* **71**, 4380-4403.

Amelin Y., Krot A. N., Hutcheon I. D. and Ulyanov A. A. (2002) Lead isotopic ages of chondrules and calcium-aluminum-rich inclusions. *Science* **297**, 1678-1683.

Birck J.-L. and Allègre C. J. (1984) Chromium isotopic anomalies in Allende refractory inclusions. *Geophys. Res. Lett.* **11**, 943-946.

Birck J.-L. and Allègre C. J., (1985) Isotopes produced by galactic cosmic rays in iron meteorites. In: Gautier, D. (Ed.), *Isotopic Ratios in the Solar System*. Centre National d' Études Spatiales (CNES), Cépadues-Editions, Toulouse, France.

Bizzarro M., Baker J. A., Haack H. and Lundgaard K. L. (2005) Rapid timescales for accretion and melting of differentiated planetesimals inferred from $^{26}$Al-$^{26}$Mg chronometry. *Astrophys. J.* **632**, L41-L44.

Bland P. A., Stadermann F. J., Floss C., Rost D., Vicenzi E. P., Kearsley A. T. and Benedix G. K. (2007) A cornucopia of presolar and early solar system materials at the micrometer size range in primitive chondrite matrix. *Meteorit. Planet. Sci.* **42**, 1417-1427.

Bose M., Floss C. and Stadermann F. J. (2010) An investigation into the origin of Fe-rich presolar silicates in Acfer 094. *Astrophys. J.* **714**, 1624-1636.
30

Table 1. Cr and O-isotopic compositions measured in $^{54}$Cr-rich grains in Orgueil residue.

| Grain | $\delta^{50}Cr$ | $\delta^{53}Cr$ | $\delta^{54}Cr$ | $\delta^{54}Cr_{corr}$ | $\delta^{17}O/^{16}O$ | $\delta^{18}O/^{16}O$ |
|---|---|---|---|---|---|---|
| 2_3 | 5±29[a] | -16±20 | 133±44 | | n.m. | n.m. |
| 3_1 | 97±39[a] | -42±25 | 206±63 | | n.m. | n.m. |
| 5_6 | 293±26[a] | -14±16 | 142±36 | | n.m. | n.m. |
| 5_7 | -1±35[a] | -18±25 | 271±58 | | n.m. | n.m. |
| 6_5 | 36±7[a] | -5±5 | 204±10 | | n.m. | n.m. |
| 8_3 | 17±10 | 3±8 | 1040±23 | >10000 | n.m. | n.m. |
| 10_8 | 32±12 | 0±9 | 376±19 | | n.m. | n.m. |
| 7_5 | 29±21 | -69±14 | 134±29 | <4300 | -70 ± 130 | 30 ± 66 |
| 7_10 | -80±90 | -5±66 | 1460±200 | 44000+-6000 | <5,000 | <500 |
| 8_1 | -36±29 | 7±20 | 949±57 | >11000 | -80 ± 130 | -50 ± 60 |

[a] $^{48}$Ti was not measured for these grains, thus no Ti interference correction was made to $\delta^{50}Cr$.
$\delta^{54}Cr_{corr}$: Corrected value after subtraction the contribution from surrounding material (section 3.4 and Fig. 6)
"n.m." means not measured; error bars are 1σ

Table 2. Cr isotopic composition of invidual chondrules in primitive CO and CR chondrites.

| | | $\varepsilon^{53}Cr$ | $\varepsilon^{54}Cr$ | Cr (%) |
|---|---|---|---|---|
| Ornans (CO3) | #1 | 0.06±0.05 | 1.11±0.1 | 0.7 |
| | #2 | 0.10±0.04 | 0.72±0.08 | 3.4 |
| | #3 | 0.00±0.04 | 0.54±0.21 | 4.1 |
| EET 92042 (CR2) | #1 | 0.15±0.03 | 1.46±0.12 | 0.2 |
| | #2 | 0.00±0.03 | 1.30±0.18 | 5.8 |
| | #3 | -0.01±0.04 | 1.25±0.10 | 6.0 |

Table 3. The relative abundances of the $^{18}$O- or $^{16}$O-rich SN silicate grains among different types of primitive materials.

| Sample | Presolar Silicates | | |
| --- | --- | --- | --- |
| | Total | #SN | SN/total (%) |
| Meteorites | | | |
| Acfer 094 | 172 | 25 | 14.5 ± 3.2 |
| ALH 77307 | 81 | 11 | 14 ± 4 |
| QUE 99177 | 65 | 5 | 8 ± 4 |
| MET 00426 | 28 | 4 | 14 ± 8 |
| Total | 346 | 45 | 13.0 ± 2.0 |
| Interplanetary Dust Particles | 32 | 12 | 38 ± 13 |
| Antarctic Micrometeorites | 19 | 7 | 37 ± 16 |
| Wild-2 | 4 | 1 | 25 ± 28 |

Data sources:
Meteorites (Bland et al., 2007; Bose et al., 2010; Floss and Stadermann, 2009; Mostefaoui and Hoppe, 2004; Nagashima et al., 2004; Nguyen and Zinner, 2004; Nguyen et al., 2007; Nguyen et al., 2010; Vollmer et al., 2008; Vollmer et al., 2009)

Interplanetary Dust Particles (Busemann et al., 2009; Floss et al., 2006; Messenger et al., 2003; Messenger et al., 2005; Messenger et al., 2010)

Antarctic Micrometeorites (Yada et al., 2008)

Wild-2 (Leitner et al., 2010; McKeegan et al., 2006; Stadermann and Floss, 2008)

**Figure Captions:**

Fig. 1. Raw data (top) and interference-corrected data (bottom) for (a) $^{50}Cr/^{52}Cr$ and (b) $^{54}Cr/^{52}Cr$ ratios for ROIs identified in NanoSIMS mapping run 7_10; $^{54}Cr$-rich grain is indicated in red. Errors are 2σ.

Fig. 2. Images of $^{54}Cr$-rich grain 7_10. a) A false-color NanoSIMS δ$^{54}Cr/^{52}Cr$ (values in ‰) map of a 20x20 μm² area containing numerous sub-micron grains reveals a highly $^{54}Cr$-enriched area. b) A composite NanoSIMS isotope image (Red=$^{52}Cr$, Green=$^{56}Fe$, Blue=δ$^{54}Cr$>1000 ‰) allows unambiguous identification of the $^{54}Cr$-rich grain. c) A scanning electron micrograph of the dotted rectangular area from (a) and (b); arrows point to the ~150 nm diameter $^{54}Cr$-rich grain. (d) A composite NanoSIMS isotope image (Red=$^{16}O$, Green=$^{27}Al^{16}O$) of the same region as Fig. 2c indicates that the $^{54}Cr$-rich grain also contains aluminum.

Fig. 3. Images of $^{54}Cr$-rich grain 8_1. a) A false-color NanoSIMS $^{54}Cr/^{52}Cr$ map of a 20x20 μm² area containing numerous sub-micron grains reveals a highly $^{54}Cr$-enriched area (δ$^{54}Cr$=949 permil). b) NanoSIMS $^{52}Cr^+$ image of same area; white contour indicates location of peak $^{54}Cr$ enrichment. c) High-magnification scanning electron micrograph of area surrounding $^{54}Cr$ enrichment. Co- alignment of the NanoSIMS and SEM images of this area indicate that the $^{54}Cr$ anomaly overlaps with three small grains (circled), two of about 100 nm in diameter sitting on top of a larger, ~200 x 300 nm grain. d) A composite Auger element map (Red=Al, Green=Cr, Blue=Fe) of area shows that the larger grain is relatively Fe-rich, while the upper of the smaller grains has significant Al contents (also seen in the NanoSIMS data).

Fig. 4. a) $\delta^{54}$Cr vs. $\delta^{53}$Cr and b) $\delta^{54}$Cr vs. $\delta^{50}$Cr for $^{54}$Cr- rich oxide grains (1 σ error) and similarly sized grains identified in the same SIMS images. $^{50}$Cr was not reliably measured in some grains (see Table 1 footnotes for details), and these grains were excluded in Fig. 4b.

Fig. 5. The O-isotopic ratios of presolar oxide and silicate grains from meteorites (triangles = this study, diamonds = previous work, (Bland et al., 2007; Busemann et al., 2009; Gyngard et al., 2010; Nguyen et al., 2007; Nguyen et al., 2010; Nittler et al., 2008; Vollmer et al., 2009) ) compared with 20$M_\odot$ SN zones of Woosley & Heger (2007) (indicated with open circles). The grey-shaded data points indicate grains from supernovae; most lie along a single line representing a mixture of material from deep $^{16}$O-rich zones with a pre-existing mixture of material from the outer He/C, He/N and H zones (black square). The black circle indicates the upper limit on O-isotopic ratios predicted for grains with $\delta^{54}$Cr>10,000 ‰.

Fig. 6. Real and simulated ion images of area 7_10. a) Observed $^{52}$Cr$^{16}$O$^-$ image acquired with a 150 nm Cs$^+$ beam. b) The observed $^{52}$Cr$^+$ image acquired with a 650 nm O$^+$ ion beam; arrow and curve indicate location of $^{54}$Cr enrichment (measured $\delta^{54}$Cr~1500 ‰). c) A simulated $^{52}$Cr$^+$ image. d) The observed $\delta^{54}$Cr image; white contours indicate larger Cr-rich grain outlines. e) A simulated $\delta^{54}$Cr image based on assumption that 100 nm diameter anomalous grain has true $\delta^{54}$Cr=44,000 ‰.

Fig. 7. The Cr-isotopic composition of grains 7_10 and 7_5 compared to predicted compositions for spallation on pure Fe (triangle) and for zones of a 20 $M_\odot$ Type II supernova (circles, (Woosley and Heger, 2007)). The star indicates the measured composition of 7_10; the solid curves are ± 2σ mixing lines, assuming that the measurement includes isotopically normal Cr from neighboring grains on the sample mount. The arrow indicates the best estimate for the true composition of grain 7_10 after correction for isotopic dilution and the shaded region represents the lower limit on the true $^{54}Cr/^{52}Cr$ ratio. Dashed curves are similar ± 2σ mixing lines for grain 7_5. a) $^{54}Cr/^{52}Cr$ versus $^{53}Cr/^{52}Cr$. b) $^{54}Cr/^{52}Cr$ versus $^{50}Cr/^{52}Cr$.

Fig. 8: Predicted correlations (solid lines) between Cr and Ti isotopic ratios based on mixing of material from the O/Ne zone of a 20 $M_\odot$ supernova (Woosley and Heger, 2007) with solar composition. Dashed lines indicate trends observed in bulk meteorites and other planetary bodies (Trinquier et al., 2009). a) $^{46}Ti/^{47}Ti$ plotted versus $^{54}Cr/^{52}Cr$; ratios are expressed as epsilon values. b) $^{50}Ti/^{47}Ti$ plotted versus $^{46}Ti/^{47}Ti$.

**Appendix:**

Fig. A1 illustrates three tests of the statistical significance of the detected $^{54}$Cr anomalies: $\delta^{54}$Cr ratios plotted against the $^{52}$Cr$^+$ secondary ion intensity, 1 sigma error on $\delta^{54}$Cr, and $^{56}$Fe/$^{52}$Cr ratio. Although most of the anomalous regions were skewed to the relatively low Cr ion intensity areas, the range of their Cr ion intensity is overlap with that defined by the isotopically normal grains, implying that the observed anomalies are unlikely artifacts associated with low Cr ion counts (Fig. A1a). In addition, the fact that most of the anomalous regions were skewed to the relatively low Cr ion intensity areas suggests that the $^{54}$Cr-rich grains are not large (μm-sized) and/or Cr is not a major element in them. Fig. A1b shows that all the anomalous regions are outside the 3σ envelope and, therefore, that their enrichments are statistically significant.

The potential interference from $^{54}$Fe was corrected assuming that the Fe in the Orgueil residue is isotopically normal and homogeneous. The fact that the anomalous regions always have relatively low $^{56}$Fe/$^{52}$Cr ratios suggests that these anomalies are not likely due to a misapplied Fe interference correction (Figure A1c). This can be tested by assuming a $^{56}$Fe/$^{52}$Cr ratio of 0.1 in the grains, since most grains have $^{56}$Fe/$^{52}$Cr ratio < 0.1. Under these circumstances, this $^{56}$Fe/$^{52}$Cr ratio, to generate a $\delta^{54}$Cr excess of 200 ‰, requires a $^{54}$Fe/$^{56}$Fe ratio of 1.89 × solar. Likewise, to generate an anomaly of 1000 ‰, a $^{54}$Fe/$^{56}$Fe ratio of 4.43 × solar is needed. Such a $^{54}$Fe/$^{56}$Fe has never been found in presolar grains of any type (Marhas et al., 2008). Thus, the observed $^{54}$Cr enrichments are almost certainly not due to $^{54}$Fe excesses in these grains.

In addition, repeat measurements for three anomalous regions focused on small subareas around the anomalous grains (6_5, 8_1 and 8_3) all confirmed the anomalies. Fig. A2 shows a repeated measurement for Run 8_3.

Figure Captions:

Fig. A1. Corrected $\delta^{54}Cr$ (in ‰) plotted against a) $^{52}Cr$ count rate, b) 1σ error on $\delta^{54}Cr$ and c) $^{56}Fe/^{52}Cr$ for $^{54}Cr$- rich oxide grains and similarly sized grains identified in the same SIMS images. The gray lines in Fig. A1b show 3 σ-envelop.

Fig. A2. $^{54}Cr$-rich anomaly in Run 8_3 (a, 25 × 25 μm raster) was confirmed and revealed to be more extreme when we re-imaged an area of 5 × 5μm around the $^{54}Cr$-rich region (b).

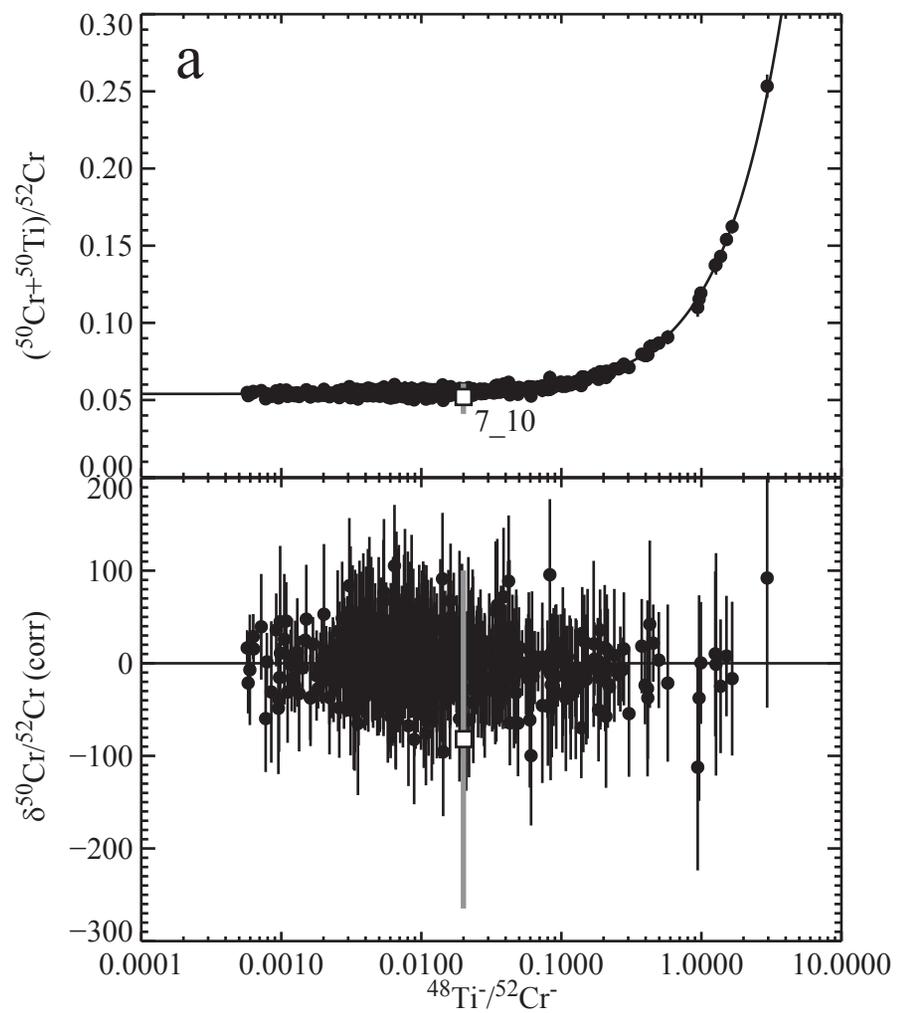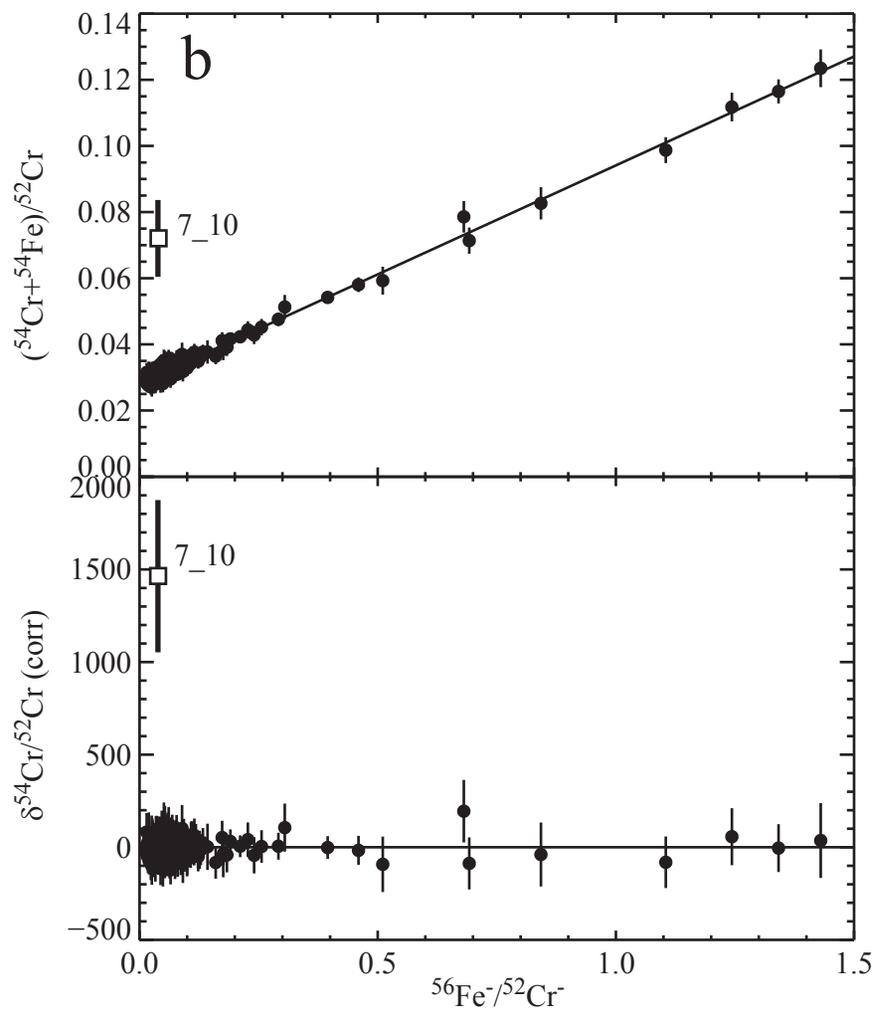

Fig. 1

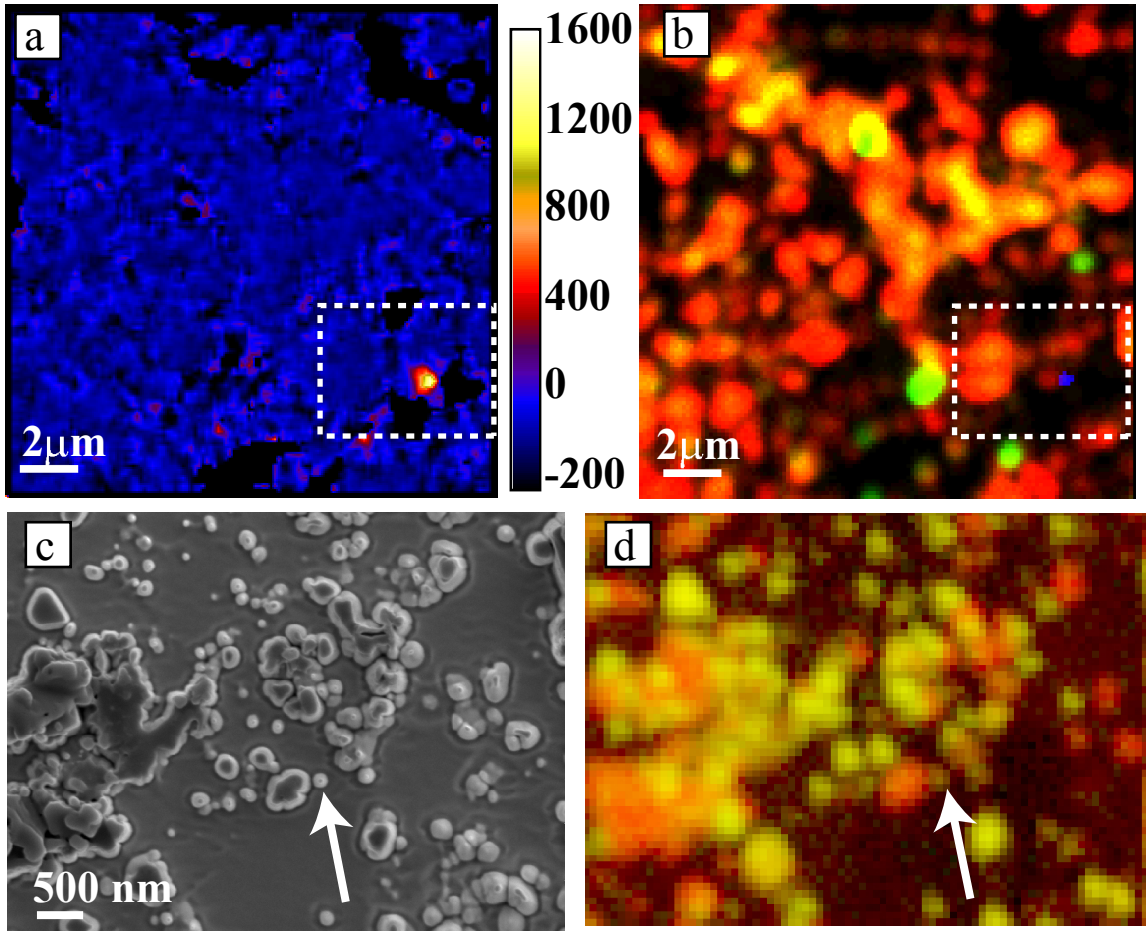

Fig. 2

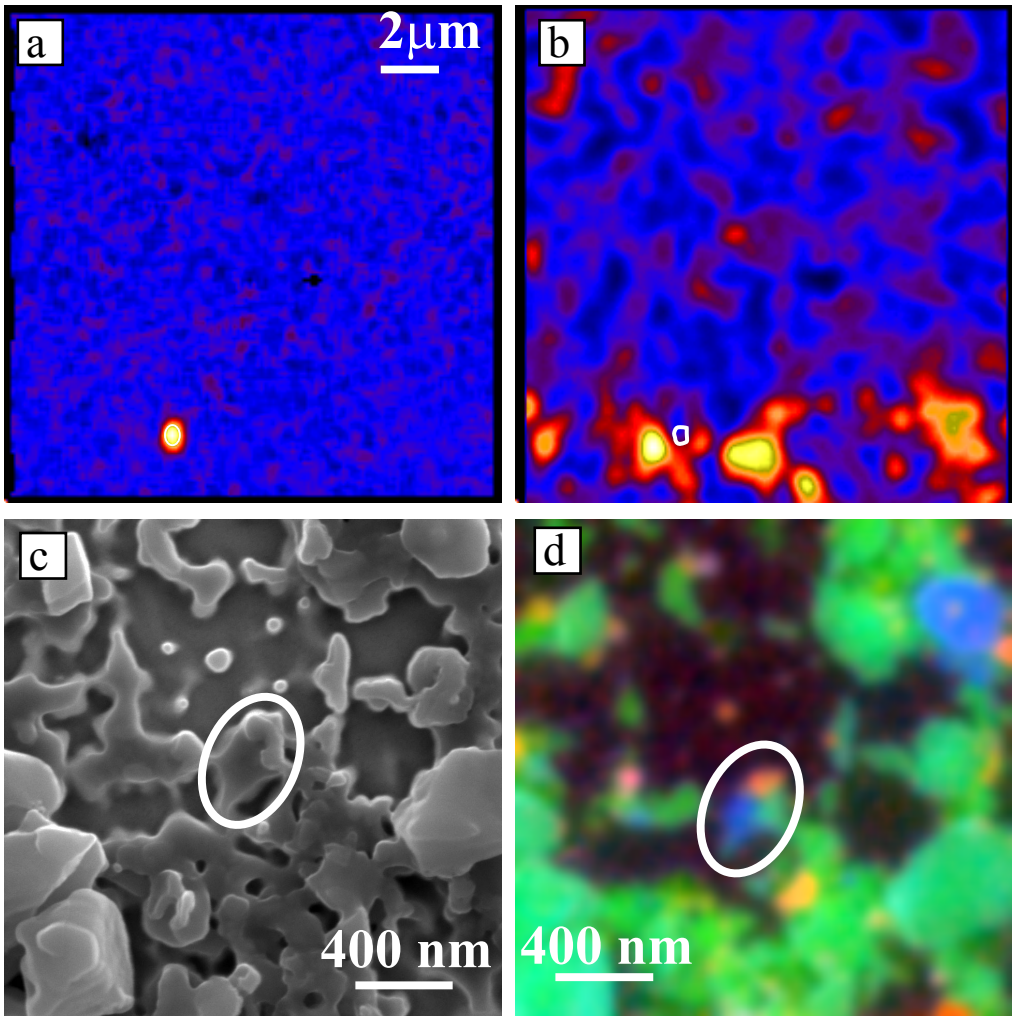

Fig. 3

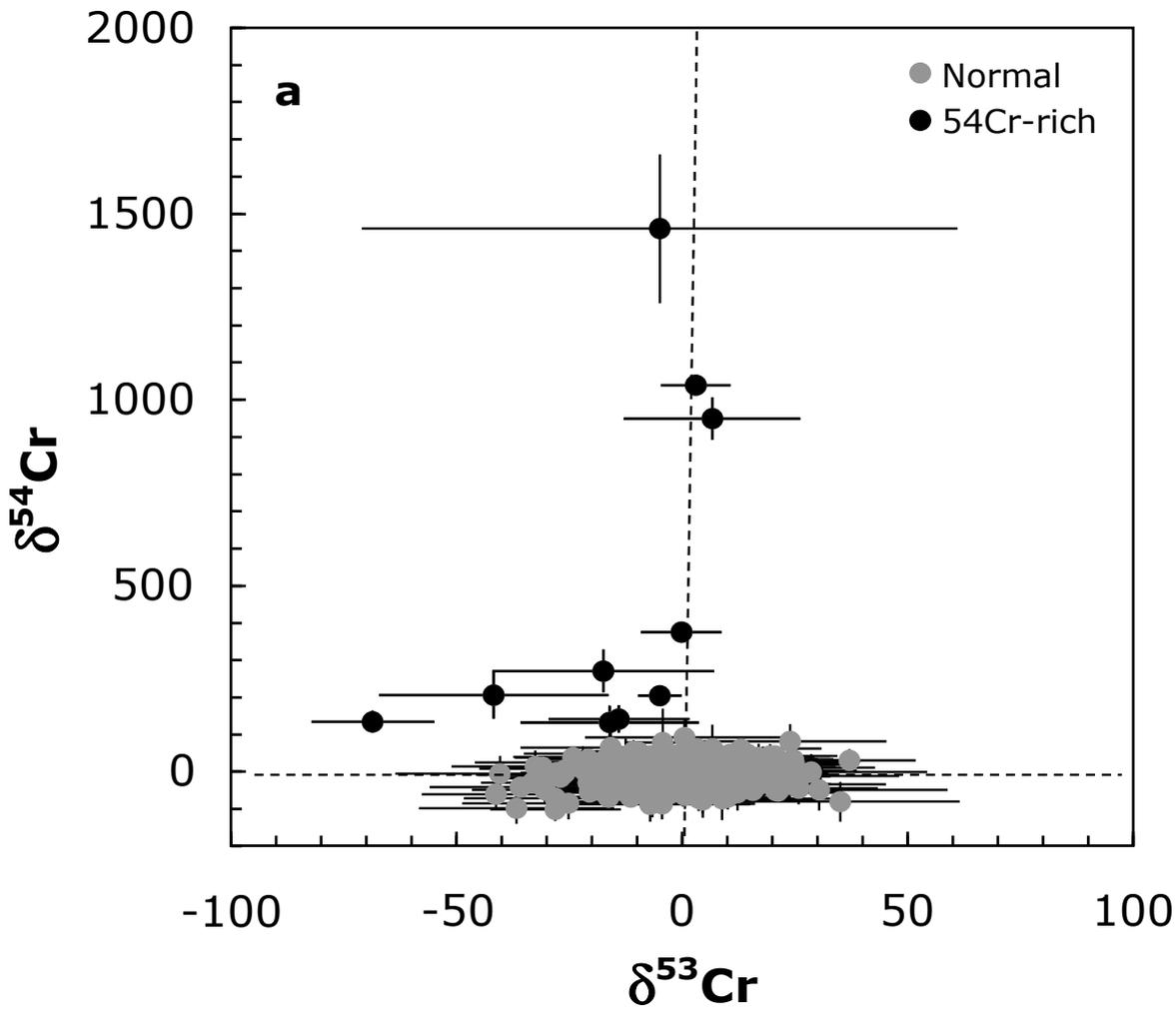

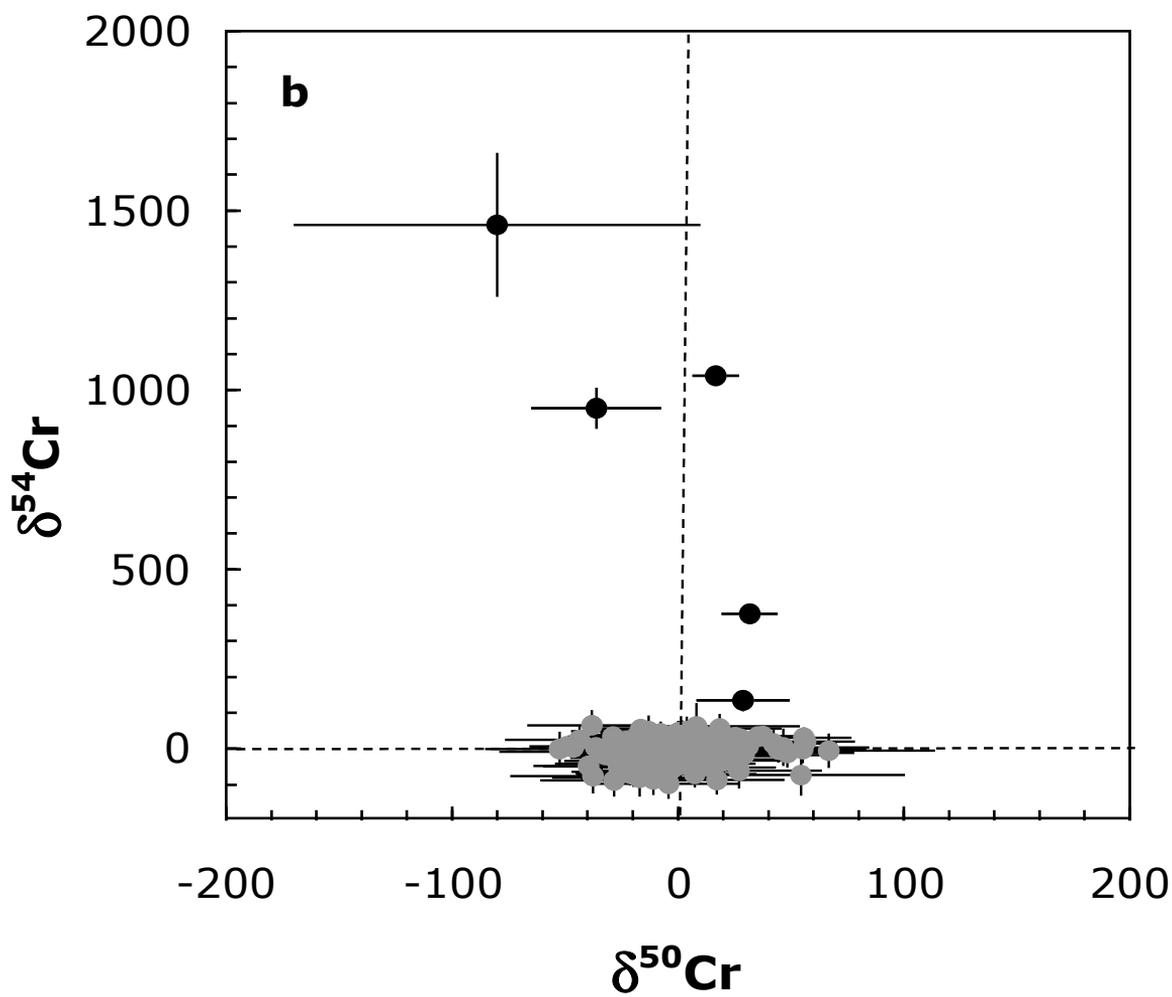

Fig. 4

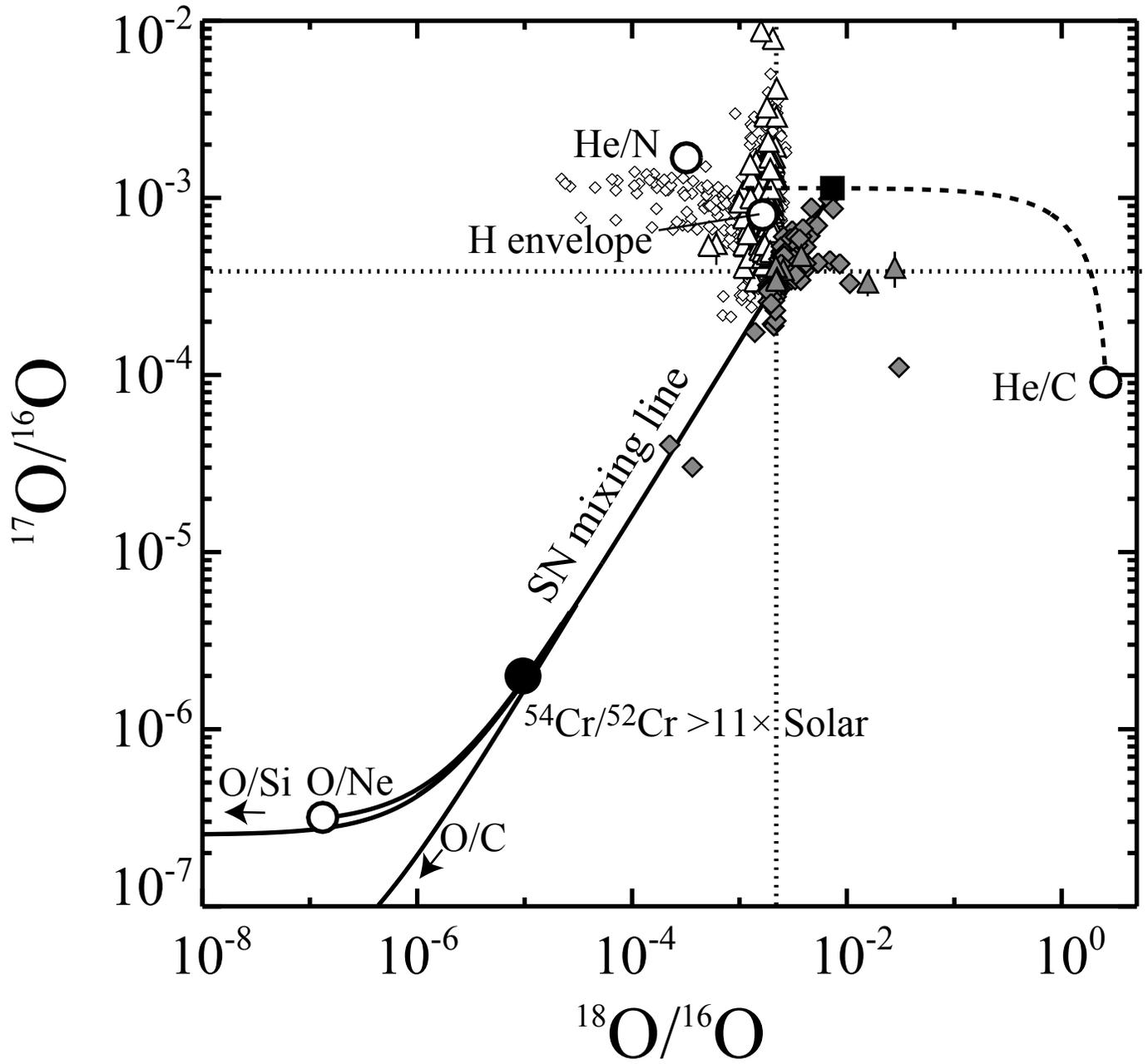

Fig. 5

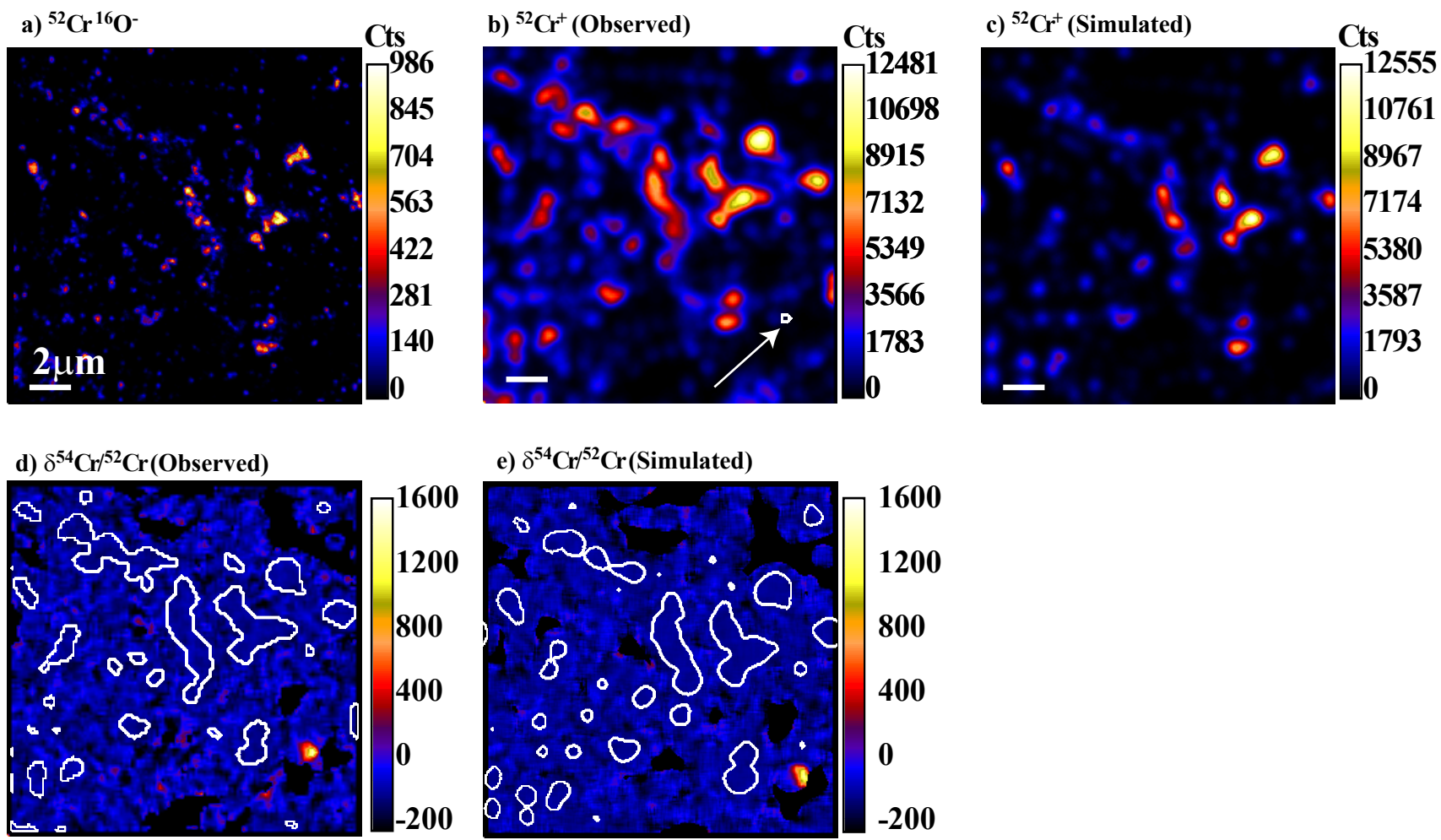

Fig. 6

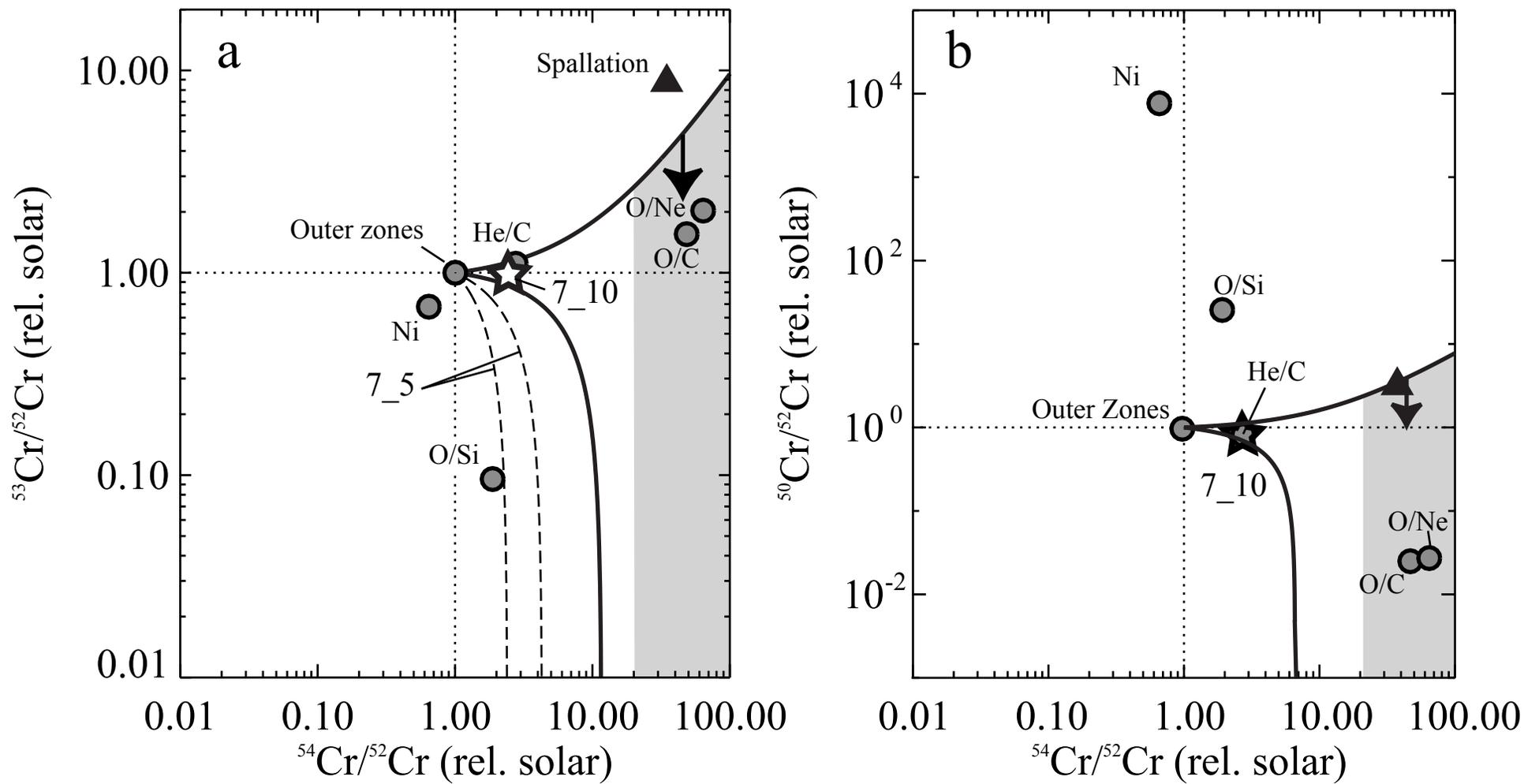

Fig. 7

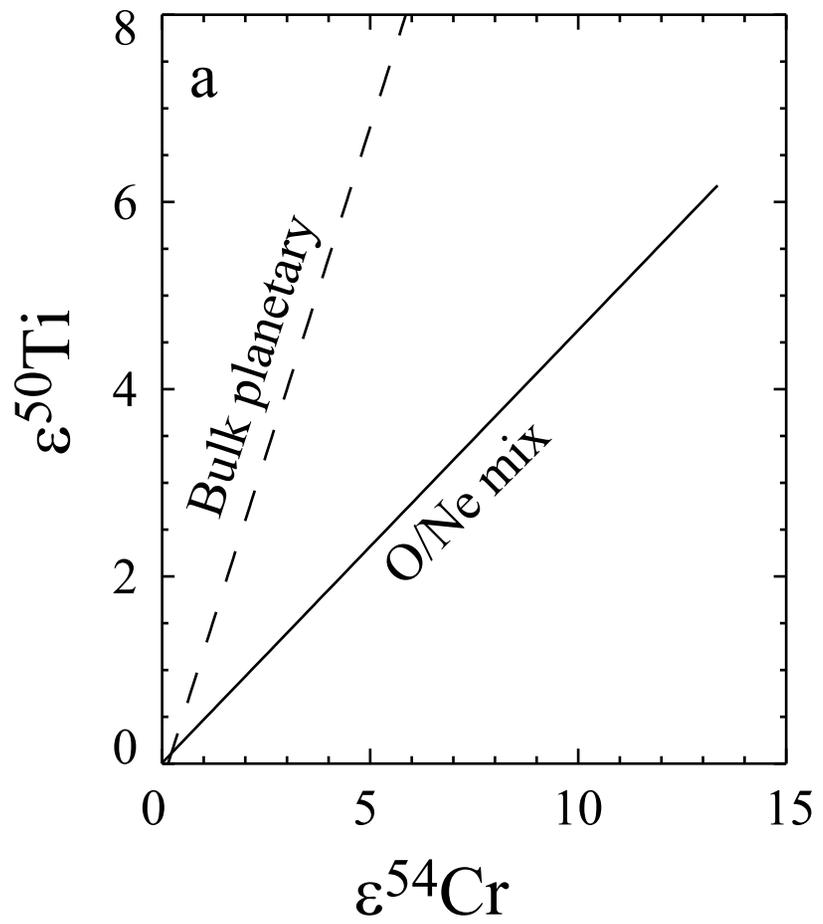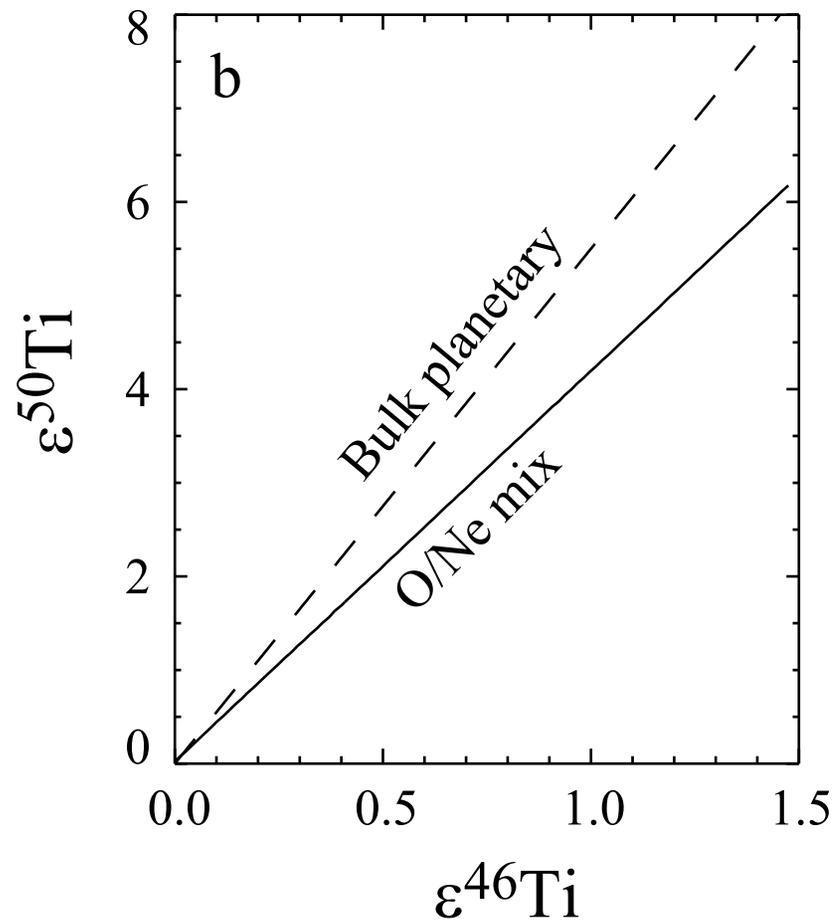

Fig. 8

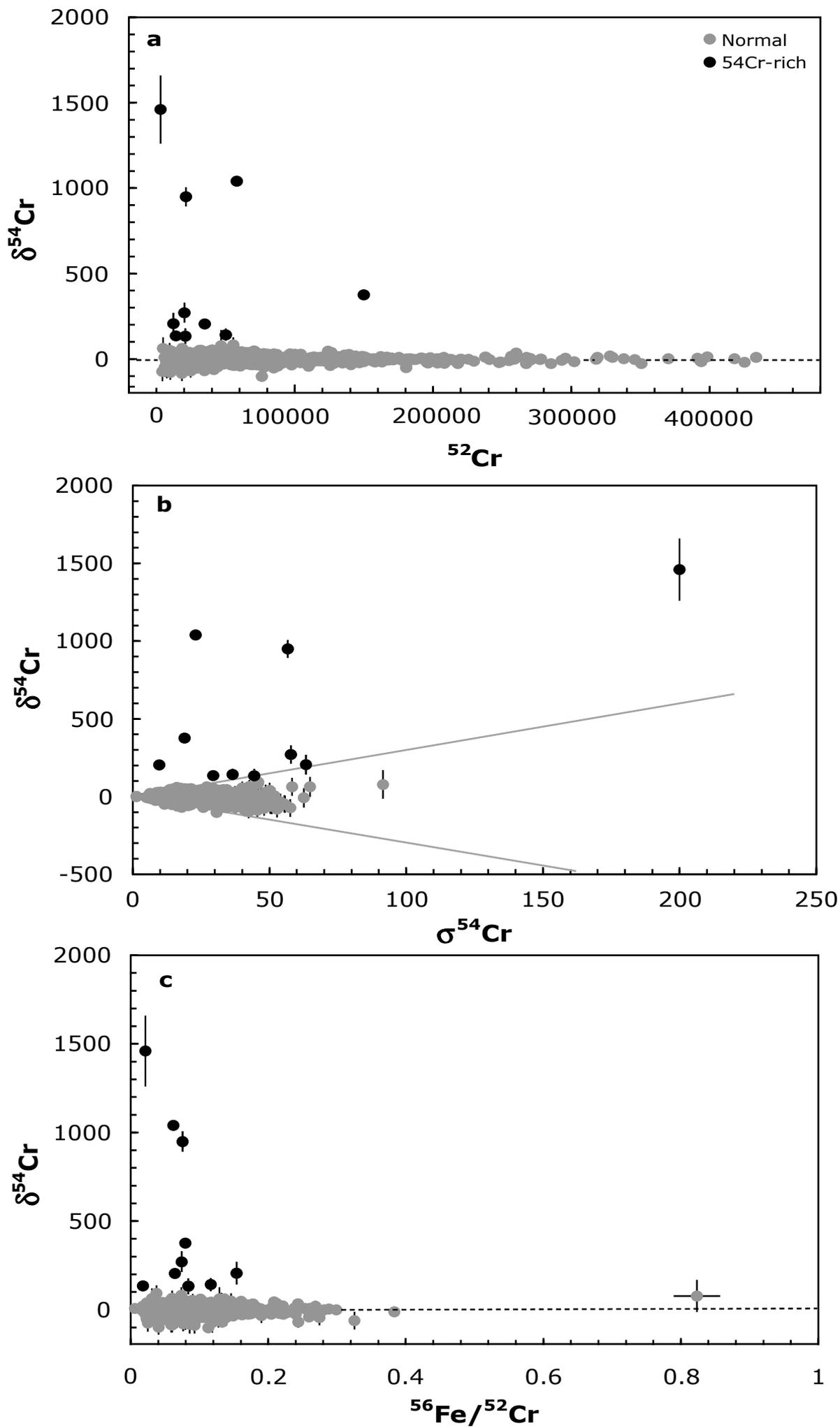

Fig. A1

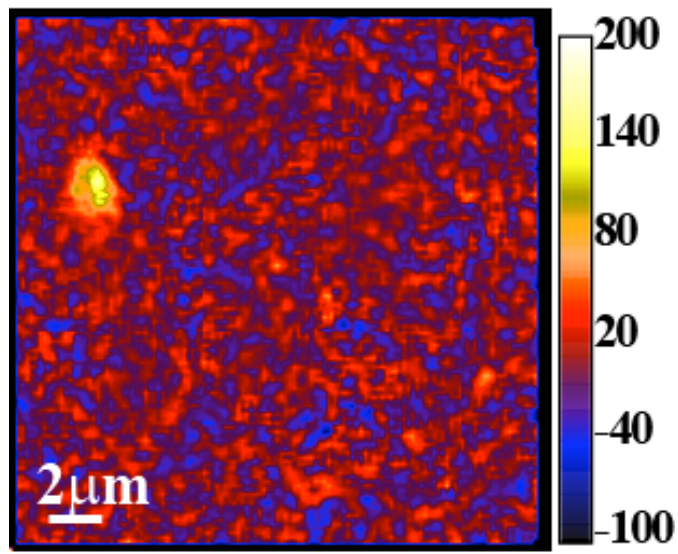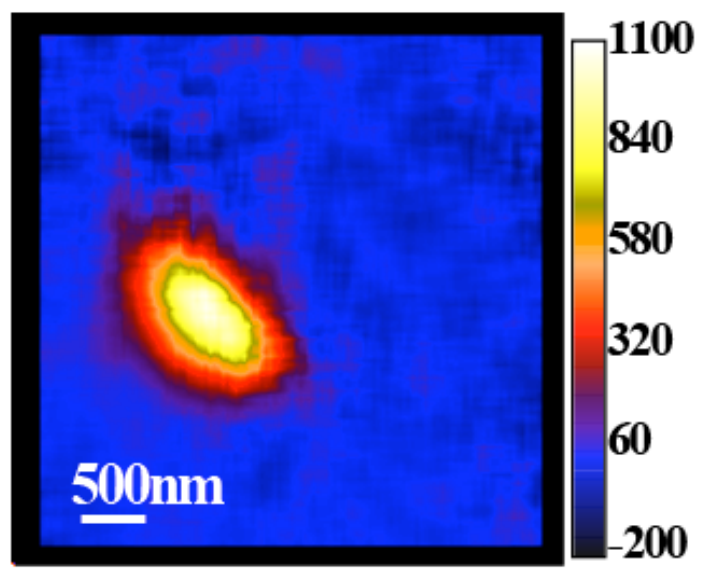

Fig. A2